\documentclass[fleqn,10pt]{wlscirep}
\usepackage[utf8]{inputenc}
\usepackage[T1]{fontenc}
\usepackage{bm}
\usepackage{xr}
\usepackage{cleveref}
\usepackage{mathtools}
\usepackage[mathscr]{euscript}
\usepackage{float}
\usepackage{url}
\usepackage{multibib}  
\newcites{main}{Reference} 

\title{Virtual Node Graph Neural Network for Full Phonon Prediction}
\author[1,2,$\dagger$]{Ryotaro Okabe}
\author[1,3,$\dagger$]{Abhijatmedhi Chotrattanapituk}
\author[1,4]{Artittaya Boonkird}
\author[5]{Nina Andrejevic}
\author[3]{Xiang Fu}
\author[3]{Tommi S. Jaakkola}
\author[6]{Qichen Song}
\author[1,4]{Thanh Nguyen}
\author[1,7]{Nathan Drucker}
\author[8]{Sai Mu}
\author[9]{Bolin Liao}
\author[10]{Yongqiang Cheng}
\author[1,4,*]{Mingda Li}
\affil[1]{Quantum Measurement Group, Massachusetts Institute of Technology, Cambridge, MA, USA}
\affil[2]{Department of Chemistry, Massachusetts Institute of Technology, Cambridge, MA, USA}
\affil[3]{Department of Electrical Engineering and Computer Science, Massachusetts Institute of Technology, Cambridge, MA, USA}
\affil[4]{Department of Nuclear Science and Engineering, Massachusetts Institute of Technology, Cambridge, MA, USA}
\affil[5]{Argonne National Laboratory, Lemont, IL, USA}
\affil[6]{Department of Chemistry and Chemical Biology, Harvard University, Cambridge, MA, USA}
\affil[7]{Applied Physics, School of Engineering and Applied Sciences, Harvard University, Cambridge, MA, USA}
\affil[8]{Department of Physics and Astronomy, University of South Carolina, Columbia, South Carolina, USA}
\affil[9]{Department of Materials, University of California, Santa Barbara, Santa Barbara, CA, USA}
\affil[10]{Chemical Spectroscopy Group, Spectroscopy Section, Neutron Scattering Division Oak Ridge National Laboratory, Oak Ridge, TN, USA}
\affil[$\dagger$]{These authors contributed equally}
\affil[*]{e-mail: mingda@mit.edu}

\begin{abstract}        
The structure-property relationship plays a central role in materials science. Understanding the structure-property relationship in solid-state materials is crucial for structure design with optimized properties. The past few years witnessed remarkable progress in correlating structures with properties in crystalline materials, such as machine learning methods and particularly graph neural networks as a natural representation of crystal structures. However, significant challenges remain, including predicting properties with complex unit cells input and material-dependent, variable-length output. Here we present the virtual node graph neural network to address the challenges. By developing three types of virtual node approaches - the vector, matrix, and momentum-dependent matrix virtual nodes, we achieve direct prediction of $\Gamma$-phonon spectra and full dispersion only using atomic coordinates as input. We validate the phonon bandstructures on various alloy systems, and further build a $\Gamma$-phonon database containing over 146,000 materials in the Materials Project.
Our work provides an avenue for rapid and high-quality prediction of phonon spectra and bandstructures in complex materials, and enables materials design with superior phonon properties for energy applications. The virtual node augmentation of graph neural networks also sheds light on designing other functional properties with a new level of flexibility. \\

\end{abstract}
\begin{document}
\flushbottom
\maketitle
\thispagestyle{empty}
\section*{Introduction} 
The structure-property relationship defines one of the most fundamental questions in materials science\cite{oganov2006crystal,le2012quantitative}. The ubiquitous presence of structure-property relationships profoundly influences almost all branches of materials sciences, such as structural materials\cite{cheng2011atomic}, energy harvesting and conversion and energy storage materials\cite{mishra2009metal, dresselhaus2007new, liu2018antiferroelectrics}, catalysts\cite{zheng2021metal} and polymers\cite{jancar2010current}, and quantum materials\cite{kumar2020topological}. However, despite its central importance to materials design, building an informative structure-property relationship can be nontrivial. On the one hand, the number of stable structures grows exponentially with unit cell size\cite{oganov2019structure}, and the structure design efforts have been largely limited to crystalline solids with relatively small unit cells. On the other hand, certain material properties are challenging to acquire due to experimental or computational complexities. \\

In the past few years, data-driven and machine-learning methods play an increasingly important role in materials science and significantly boost the research on building structure-property relationships\cite{zhu2021charting, dunn2020benchmarking, peng2022human}. Complex structures such as porous materials\cite{altintas2021machine, schwalbe2021priori}, nanoalloys\cite{yao2022high, hart2021machine}, and grain boundaries\cite{wagih2020learning} are becoming more feasible to handle, and properties ranging from mechanical strength to quantum ordering can be learned with increased confidence\cite{guo2021artificial, stanev2021artificial}. One particular powerful approach is the graph neural networks (GNNs)\cite{fung2021benchmarking}. By representing atoms as graph nodes and interatomic bonds as graph edges, GNNs provide a natural representation of molecules and materials. For crystalline solids, crystallographic symmetry offers a further boost on the GNN performance, with a few symmetry-augmented GNNs being proposed \cite{xie2018crystal, Thomas2018TFN, geiger2022e3nn}. A few fundamental challenges still exist. For one thing, many materials properties are not naturally represented as a weighted aggregation of each atom in real space, such as reciprocal and energy space properties. For another thing, the output property length is usually fixed, like the heat capacity\cite{delaire2009phonon} as a single scalar. In contrast, many materials' properties have unique degrees of dimensions, such as the number of electronic and phononic bands\cite{baroni:phonon:2001}, frequency ranges with optical responses, and the features of magnetic structures like propagation vectors. \\

In this work, we propose Virtual Node Graph Neural Network (VGNN) as a generically applicable approach to augment GNN. In contrast to symmetry-augmented GNN which focuses on reducing the input data volume, VGNN focuses on handling the output properties with variable or even arbitrary dimensions. We study materials' phonon spectra and dispersion relations, given that phonons bands are challenging to compute or measure with high computational cost and limited experimental resources. By using the phonon spectra as examples, we present three versions of VGNN: the vector virtual nodes (VVN), the matrix virtual nodes (MVN), and the momentum-dependent matrix virtual nodes ($k$-MVN). All three VGNN models take atomic structures as input without prior knowledge of interatomic forces. The VVN is the simplest VGNN that takes in a crystal structure with $m$ atoms and outputs $3m$ branches $\Gamma$-phonon energies. The MVN is a more involved VGNN that shows higher accuracy for complex materials with slightly higher computational cost. Finally, the $k$-MVN is a VGNN that can predict full phonon band structure at arbitrary $k$ points in the Brillouin zone. To achieve so, the crystal graphs contain "virtual-dynamical matrices", which are matrix structures that resemble phonon dynamical matrices\cite{kong2011phonon}. Instead of performing direct \textit{ab initio} calculations on each material, all matrix elements are learned from the neural network optimization process using training data comprised of all other materials.  Our work offers an efficient technique that can compute zone-center phonon energies and full phonon band structures directly from atomic structures in complex materials and enables phonon property optimization within a larger structure design space. The prediction methods has enabled us to acquire relevant information of materials such as group velocities, heat capacities, density of states as by-products. Meanwhile, the virtual node structures also shed light on future flexible GNN design, that to put intermediate crucial quantities (e.g. dynamical matrix) as key learning parameters without having to put target properties (e.g. phonon band structures) as output. 

\begin{figure}[ht!]
\includegraphics[width=\textwidth]{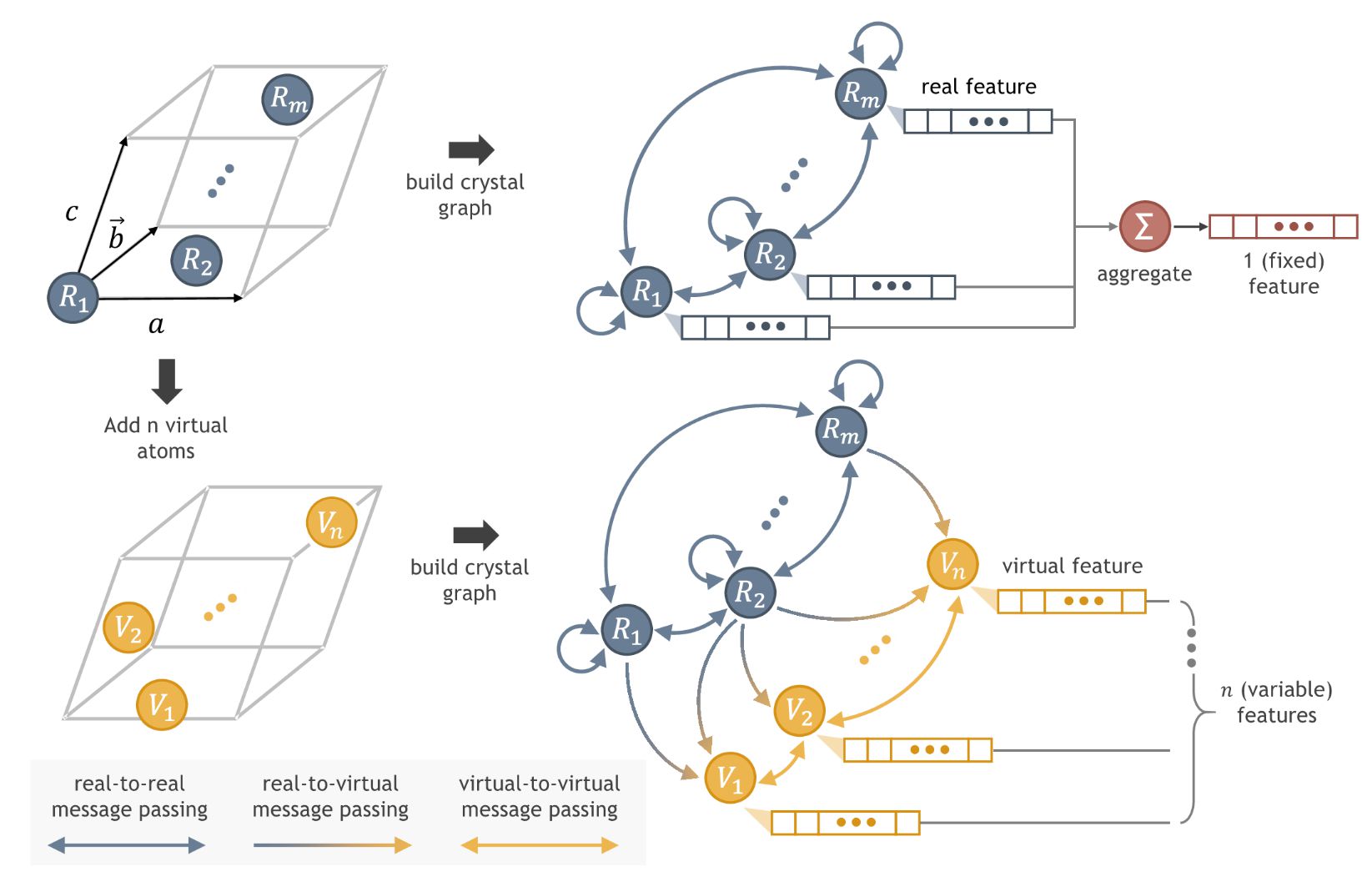}
\caption{\textbf{Overview of virtual node graph neural network (VGNN).} 
\textbf{a}. Atomic structure of a crystalline material with $m$ atoms per unit cell. 
\textbf{b}. A GNN converts the atomic structures into a crystal graph. After layers of graph convolutions (omitted for simplicity), the final node features are aggregated into a single fixed-sized output feature.
\textbf{c}. A flexible of $n$ virtual atoms are added into the crystal structure. \textbf{d}. After forming the crystal graph with both real and virtual nodes, the flexibility of virtual nodes enables the choices of output not necessarily from real-node aggregation but can have variable length and in different spaces.}
\label{fig_overview}
\end{figure}

\section*{Results}
\textbf{Virtual node augmentation for graph neural networks.} Figure \ref{fig_overview} gives an overview of the VGNN method as a generic approach to augment GNN. For a crystal with $m$ atoms per unit cell (Figure \ref{fig_overview}a), a typical GNN model converts the crystal into a crystal graph, where each graph node represents an atom, and each graph edge represents the interatomic bonding as shown in Figure \ref{fig_overview}b. The node features associated with each atomic node (gray arrays in Figure \ref{fig_overview}b) are updated by neighborhood nodes and edges connecting the nodes (gray arrows in Figure \ref{fig_overview}b). After iterative layers of graph convolutions, $m$ final-layer node features are obtained that represent the atomic features from each of the $m$ atoms. The final graph output can be obtained by aggregating the final-layer node features into one fixed-sized output. 

Figures \ref{fig_overview}c,d describe the general idea of VGNN that endows a GNN with greater flexibility for prediction. On top of the conventional, real-node GNN, virtual atoms are added into crystal (yellow nodes in Figure \ref{fig_overview}c), which become the virtual nodes in the corresponding GNN (yellow nodes in Figure \ref{fig_overview}d). As Figure \ref{fig_overview}d illustrates, just like the bi-directional message passing between real atomic nodes (double-arrow gray lines), the message passing (double-arrow yellow lines) between virtual nodes is also bi-directional. On the other hand, to preserve the structure of the conventional GNN, the messages from real nodes to virtual (single-arrow gray-to-yellow gradient lines) are uni-directional. Given the flexibility of the choice of the virtual nodes, a VGNN gains huge flexibility to predict materials-dependent outputs with arbitrary lengths and in spaces. We will introduce three VGNN methods for phonon prediction with increased levels of predictive power and complexity. \\

\begin{figure}[ht!]
\includegraphics[width=\textwidth]{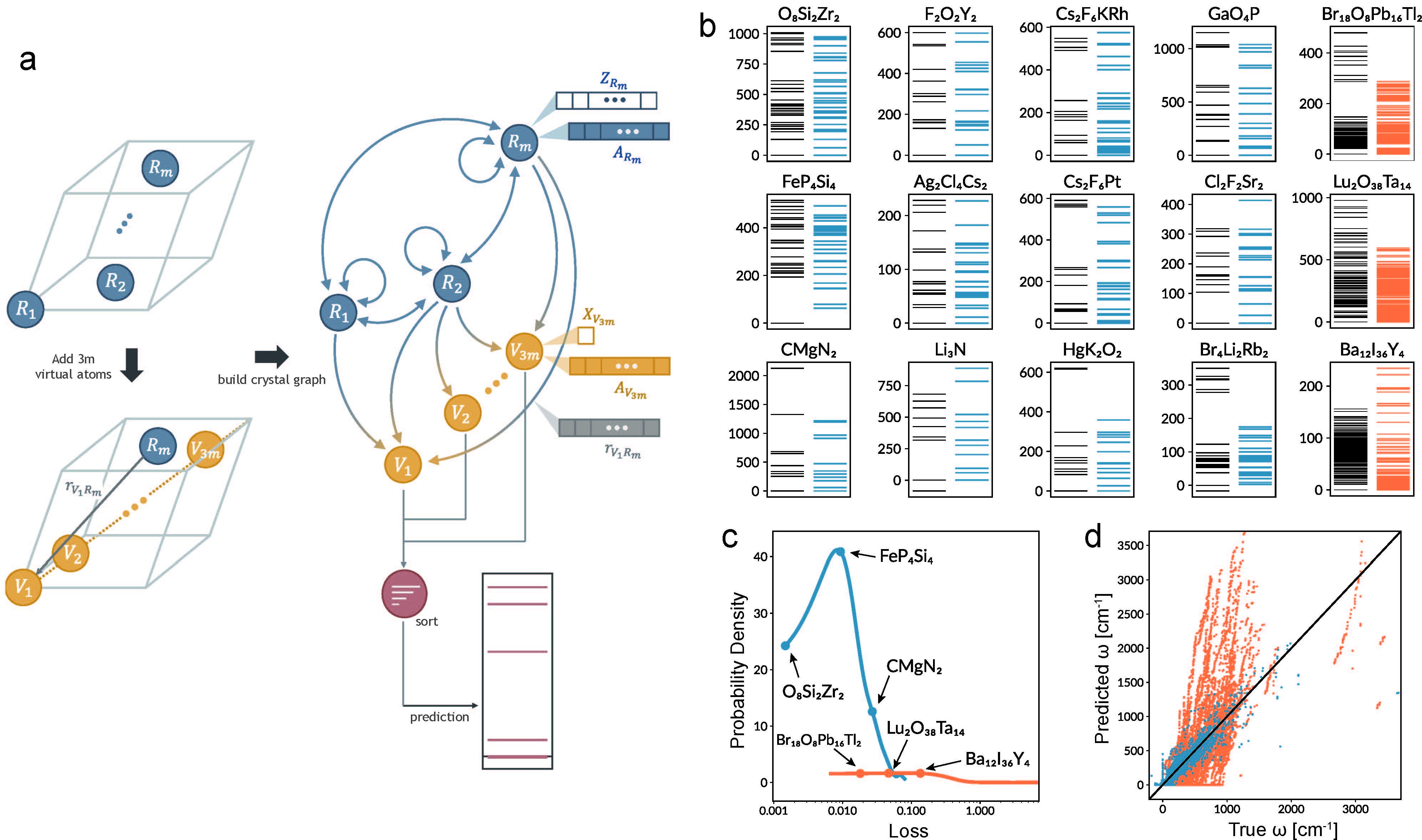}
\caption{\textbf{The vector virtual node (VVN) method to predict $\Gamma$-point phonons.}
\textbf{a}. Schematic of VVN model construction and prediction. For material with $m$ atoms per unit cell, $3m$ Virtual nodes are augmented along the diagonal vector $\vec{v} = \vec{a} + \vec{b} + \vec{c}$ of the unit cell. We embedded the components of the crystal when building the GNN model. For instance, atomic numbers of the $m^\text{th}$ real atom ($A_{R_m}$) and that of the $3m^\text{th}$ virtual atom ($A_{V_{3m}}$) are embedded as the attributes of each nodes. The atomic mass of $m^\text{th}$ real atom ($Z_{R_m}$) is set as the initial feature of that node. The relative position of the node $V_1$ with respect to $R_m$ is $\vec{r}_{V_{1}R_{m}}$, which is used to embed the edge attribute between the two nodes. The model predicts $\Gamma$-phonon spectra by sorting the scalar output features from virtual nodes.
\textbf{b}. Spectral prediction samples in the test set within each error tertile compared with ground truth (black): Test from the same database as the training set (blue), and a different database containing complex materials (orange).
\textbf{c-d}. Evaluation of the test accuracy through the distribution of  loss function and correlation plot between ground-truth and predicted average phonon frequencies, respectively. The heavy distribution at low loss regime of the distribution plot and the agreement along the diagonal line of the correlation plot for the test set (blue) indicates a high-quality phonon prediction at least for relatively simple materials with the number of atoms per unit cell $m\leq24$. The loss becomes higher with reduced performance for complex materials (orange).}
\label{fig_diagonal}
\end{figure}

\noindent\textbf{Vector virtual nodes for $\Gamma$-phonon prediction.} As illustrated in Figure \ref{fig_overview}, VGNN makes it possible to adjust output dimension based on input information with flexibility. We first introduce the vector virtual node (VVN) method, which is the simplest approach to acquire $3m$ phonon branches when inputting a crystal with $m$ atoms per unit cell. (See \hyperref[Methods]{Methods} for more detail) Figure \ref{fig_diagonal} shows the VVN approach to predict $\Gamma$-phonon spectra. Since the virtual nodes do not pass information to real nodes, there is additional flexibility in choosing the position of the virtual node. Without loss of generality, we assign the position of the virtual nodes evenly spaced along the diagonal line of the unit cell. The crystal graph is constructed with virtual and real nodes (Figure \ref{fig_diagonal}a). After updating node features in each convolution layer, the feature vectors pass a linear layer so that virtual node features $V_{i}, i\in[1, 3m]$ are converted to $3m$ scalars, which represent the predicted $\Gamma$-phonon energies. Throughout this work, the GNN part is implemented through the Euclidean neural networks\cite{geiger2022e3nn} that are aware of the crystallographic symmetry. Data preparation, neural network architectures, and optimizations are described in Supplementary Information I-III. 

The main results using the VVN for $\Gamma$-phonon prediction are shown in Figure \ref{fig_diagonal}b. The three-row spectral comparison plots are randomly selected samples from the test set within each error tertile (top-to-bottom rows are top-to-bottom performance tertiles, respectively). The first four columns are taken from the same database as the training set from high-quality density-functional perturbation theory (DFPT) calculations \cite{petretto2018high}, and the fifth column contains additional test examples with much larger unit cells from a frozen phonon database\cite{togo2015kyoto}. It is worthwhile mentioning that although for very complicated materials, the VNN-predicted phonons tend to have higher frequencies (e.g., third row, fourth column of Ba$_{12}$I$_{36}$Y$_4$), the resulted phonon density-of-states over the entire Brillouin zone can still be largely comparable. 
However, the prediction loss becomes larger and distributed broader as the input materials are more complicated (Figure \ref{fig_diagonal}c). From the correlation plot of predicted and ground-truth phonon frequencies (Figure \ref{fig_diagonal}d), most data points are along the diagonal line, indicating good prediction between VNN prediction and ground-truth from DFPT calculations with the number of atoms per unit cell $m\leq24$ (blue dots). For complex materials, the correlation performance could be degraded (orange dots). More test results are shown in Supplementary Information IV.\\

\begin{figure}[ht!]
\includegraphics[width=\textwidth]{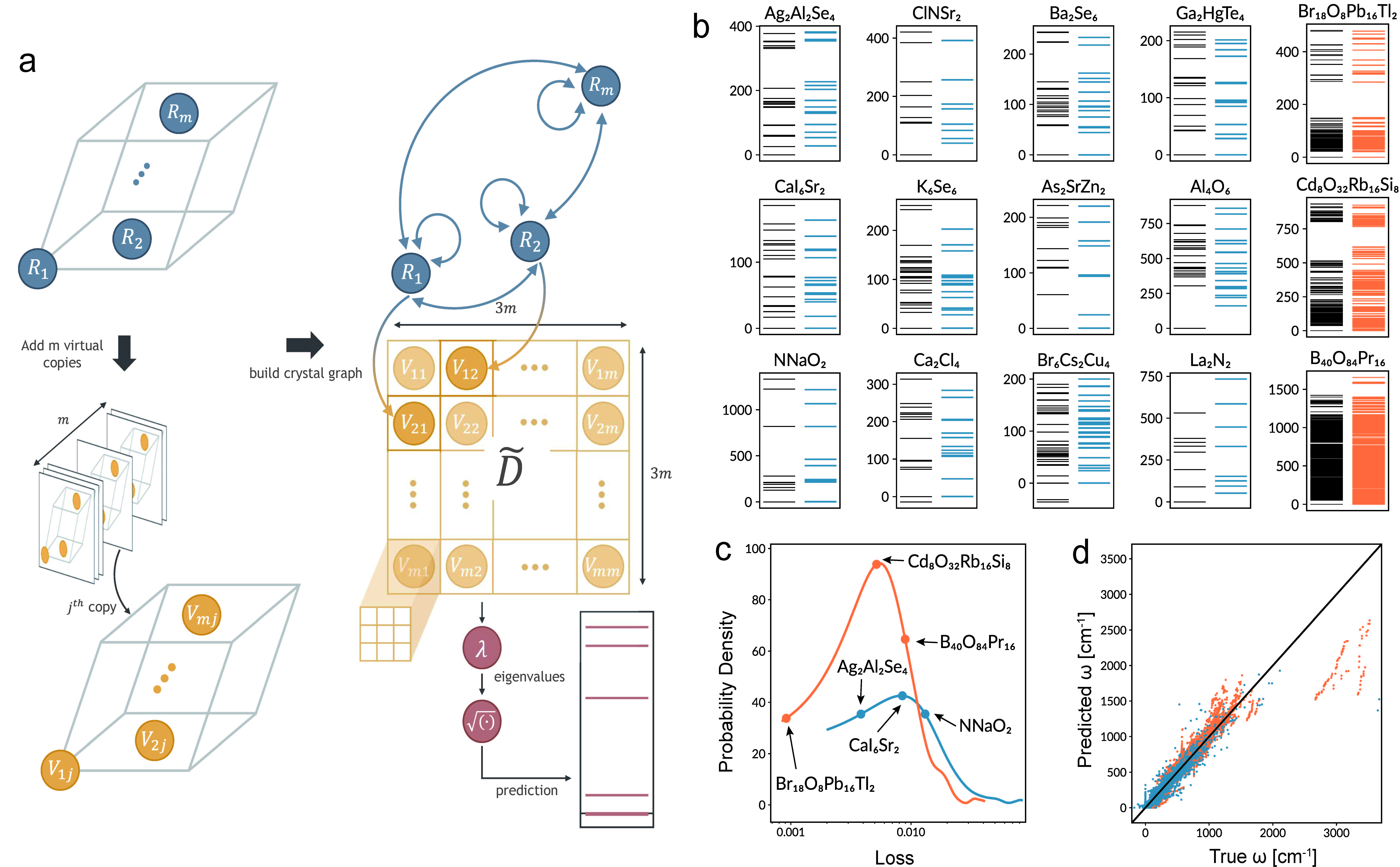}
\caption{\textbf{The matrix virtual node (MVN) method to predict $\Gamma$-point phonons.}
\textbf{a.} Augment $m^2$ virtual nodes as $m$ sets of virtual crystals (left) and the message passing scheme and post-processing of virtual node features (right). The legends are the same as Figure \ref{fig_diagonal}. In contrast to VVN, where each node $V_j$ is a scalar, here, each node $V_{ij}$ is a $3 \times 3$ matrix. The phonon spectra in MVN are obtained by solving the eigenproblems instead of direct output, as done in VVN. 
\textbf{b.} Selected test examples within each error tertile. Tests from the same dataset as the training set and additional tests containing complex materials are predicted in blue and orange, respectively. 
\textbf{c.} Comparison of prediction loss distribution with several examples of materials.  
\textbf{d.} The correlation plots of average phonon frequencies with the graph $y=x$ as reference. Better performance for MVN is achieved than VVN for complex materials (orange color), which can be seen from both the loss distribution and the average phonon frequencies.}
\label{fig_dynamic}
\end{figure}

\noindent\textbf{Matrix virtual nodes for $\Gamma$-phonon with enhanced performance.} In this section, we introduce another type of virtual nodes approach, the matrix virtual nodes (MVN). The MVN approach performs better $\Gamma$-phonon prediction than VVN, especially for complex materials, with a slightly higher computational cost. Moreover, the structure of MVN lays the groundwork for the full phonon band structures to be discussed in the next section. In MVN, $m$ copies of virtual crystals are generated for material with $m$ atoms per unit cell, and each copy  contains $m$ virtual nodes that share the same crystal structure as the real crystal (Figure \ref{fig_dynamic}a). This results in a total of $m^2$ virtual nodes $V_{ij}$, $i,j\in[1, m]$ with more involved node connectivity. (See \hyperref[Methods]{Methods} for more detail). 

With this graph construction scheme, after the neural network training, the virtual nodes $V_{ij}$ would capture the essence of the connection between $R_i$, and $R_j$. Hence, after the message passes in each convolutional layer, each virtual node feature is further converted into a three-by-three matrix. Each of $V_{ij}$ is assembled to form $(i, j)$ block of a supermatrix $\Tilde{D}$ of shape $(3m, 3m)$. Given the structural similarity of this matrix and the dynamical matrix expressed in Equation (\ref{eq_dynamical}) with $\vec{k}=0$, we predict $\Gamma$-point phonon energies by solving for $3m$ eigenvalues of the matrix $\Tilde{D}$. It is still worthwhile mentioning that although the matrix shares a similar feature with the dynamical matrix, the matrix elements are learned from neural network training and are not necessarily the matrix elements from the real dynamical matrix. An intuitive comparison is that the edge of GNN does not necessarily reflect true chemical bonding, but is more like an atomic neighbor connection.

The predicted phonons using MVN are summarized in Figure \ref{fig_dynamic}b, which shares the same structure with Figure \ref{fig_diagonal}b as error tertile plots from the high-quality DFPT database (blue) and database for complex materials (orange). MVN shows comparable performance with VVN for simple materials (blue curves in Figure \ref{fig_diagonal}c and Figure \ref{fig_dynamic}c), but shows significant performance improvement for complex materials. The prediction loss distribution of MVN shows a heavier distribution toward a lower loss regime compared to VVN (orange curves in Figure \ref{fig_diagonal}c and Figure \ref{fig_dynamic}c), and the average phonon frequencies in the correlation plot align better toward ground truth (orange dots in Figure \ref{fig_diagonal}d and Figure \ref{fig_dynamic}d). More results and correlation plots are shown in Supplementary Information IV. \\

\begin{figure}[ht!]
\includegraphics[width=\textwidth]{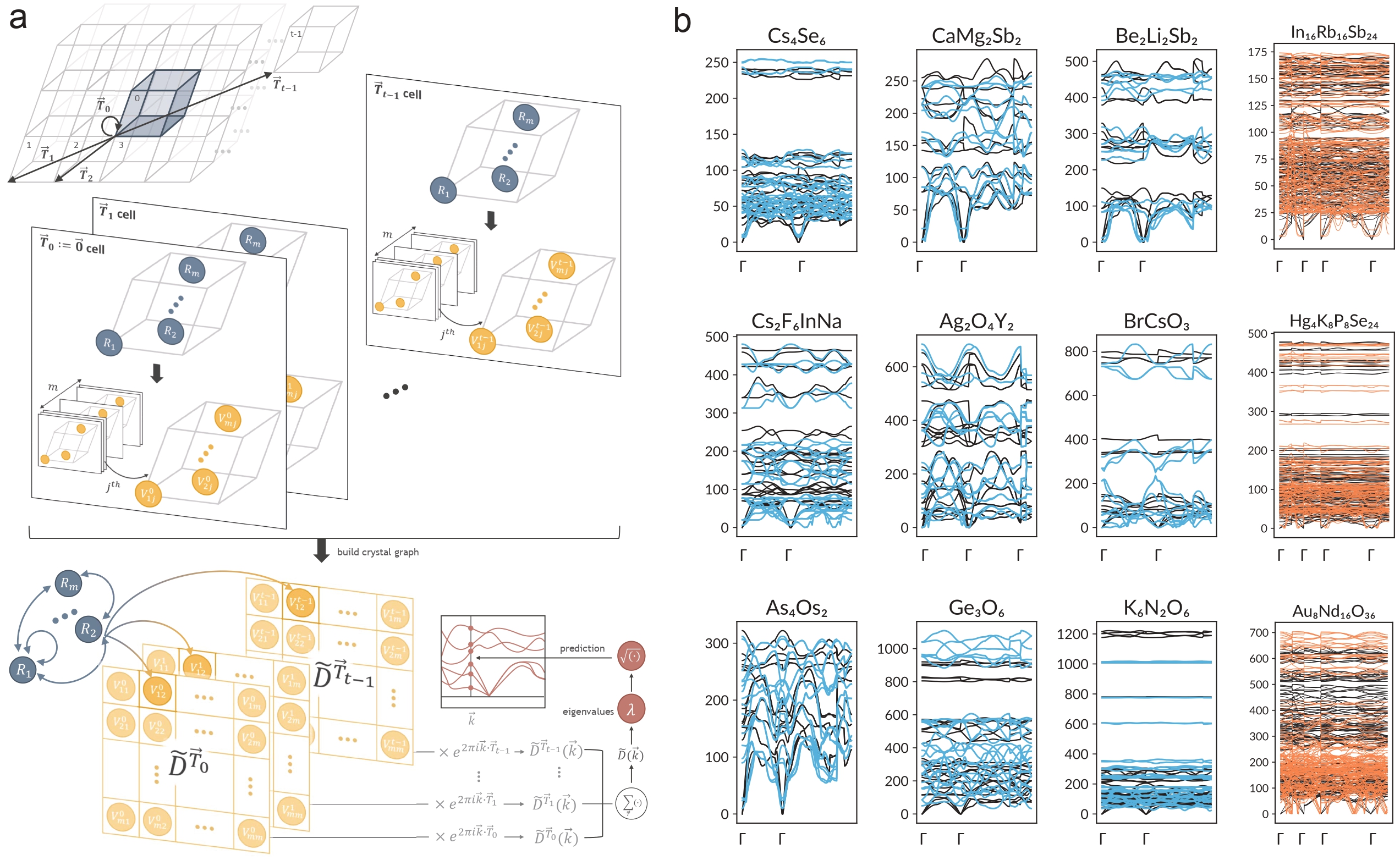}
\caption{\textbf{The momentum-dependent matrix virtual nodes ($k$-MVN) to predict full phonon band structures.}           
\textbf{a.} (Top) Augment $m^2$ virtual nodes for each translation vector $\vec{T}$, and with a total $t$ neighboring unit cells, a total $tm^2$ virtual nodes are generated. (Bottom) By multiplying a phase factor by each translated unit cell, a full virtual dynamical matrix can be constructed. \textbf{b.} Selected examples in the test set within each error tertile, for the high-quality DFPT database (blue) and additional complex materials test (orange). $\Gamma$-point positions are labeled for each spectrum. 
}
\label{fig_bands}
\end{figure}

\noindent\textbf{Momentum-dependent matrix virtual nodes for predicting full phonon band structures.}
The structure of MVN inspires us to take one step further and construct full momentum-dependent virtual dynamical matrices by taking into account the unit cell translation, termed momentum-dependent matrix virtual nodes ($k$-MVN). We construct virtual-dynamical matrices following Equation  (\ref{eq_dynamical}). In contrast to the MVN, which focuses on $\Gamma$-point phonons by taking $\vec{k}=0$, here in $k$-MVN, we include the phase factor $e^{i\vec{k}\cdot\vec{T}}$ when defining the virtual dynamical matrices, where $\vec{T}$ is the relative unit-cell translation of a neighboring unit cell origin relative to the chosen reference unit cell $\vec{T_0}$ (Figure \ref{fig_bands}a). If a total number of $t$ neighboring unit cells are included, each with translation $\vec{T_h}, h\in[0, t-1]$ (reference cell included), then a total $t$ copies of MVN-type virtual nodes matrices will be generated, with a total number of $tm^2$ virtual nodes $V^{h}_{ij},h\in[0,t-1], i,j\in[1,m]$ in $k$-MVN. To obtain the phonon band structure, each set of virtual nodes at a given $\vec{T_h}$ needs to multiply by the phase factor $e^{i\vec{k}\cdot\vec{T_h}}$, and all virtual nodes at each $\vec{T_h}$ are summed in Equation \ref{eq_dynamical_sim}. Thanks to the graph connectivity within the cutoff radius (see \hyperref[Methods]{Methods}), only a small number of $t$ is needed as long as crystal graph connectivity can be maintained. In practice, $t$ is materials dependent, and $t=27$ (nearest neighbor unit cells) is sufficient for many materials and does not need to go beyond $t=125$ (next-nearest neighbor unit cells) in all cases. Intuitively, such a supercell approach resembles the \textit{ab initio} band structure calculations with frozen phonons. To facilitate the training, phonons from selected high-symmetry points are included in the training data, without the need to use full phonon energies in the entire Brillouin zone. This significantly facilitates the training process while maintaining accuracy. More details are discussed in \hyperref[Methods]{Methods} and Supplementary Information V.   

Figure \ref{fig_bands}b shows the prediction results of phonon band structures. Here 12 materials are selected from the same dataset for training (blue color) and the additional dataset for complex materials (orange color). Despite the complexity of a generic phonon band structure, the $k$-MVN model could predict the positions and the shapes of the phonon bands, such as gaps between different optical branches. The dispersion relations of the acoustic phonons are also well generated around the $\Gamma$-points on the left three columns, even though we do not enforce that acoustic $\Gamma$-phonons have to be gapless with zero-energy known as acoustic sum rule\cite{sham1969electronic}. This may enable the prediction of crystal stability for future works. While there are risks that prediction performance could be degraded for the phonon bands of at higher frequencies, most of the predicted phonons follow the references, including the complex materials with more than 40 atoms per unit cell. 

\section*{Discussion}
We demonstrate the prediction of phonons directly from the materials’ atomic coordinates, using three different types of virtual nodes – the VVN, the MVN, and the $k$-MVN – to augment the symmetry-aware Euclidean neural networks. The comparison between the three virtual node approaches is summarized in Table  \ref{table_phonon_methods}. VVN directly acquires the phonon spectra from the virtual nodes. The assignment of $3m$ virtual nodes ensures that the output phonon band number is always $3m$ for a crystal with a primitive unit cell containing $m$ atoms. In MVN, instead of computing phonon energies directly, a virtual dynamical matrix (VDM) is constructed first, from which the phonon energies are solved as an eigenvalue problem. This step is crucial to gain robustness for complex materials prediction since intermediate quantities like force constants and dynamical matrices are considered more “fundamental” than final phonon energies to reflect the interatomic interactions. The $k$-MVN goes one step further, using the unit-cell translations to generate the momentum dependence that could be used to obtain the full phonon band structure.\\

Today, the \textit{ab initio} calculations like frozen-phonon and DFPT remain the most accurate methods for phonon calculations. Even so, since the VGNN-based phonon calculation skips the direct calculation of the material-by-material dynamical matrix, it shows significantly faster computation speed while maintaining reasonable accuracy. Additional tests on SiGe alloys, FeCoNi alloys, and other high-energy alloys are performed, which agree well with existing literature (Supplementary Information VI). Finally, by using MVN, we build a database containing the $\Gamma$-phonon spectra for over 140,000 materials listed in Materials Project (Supplementary Data and Supplementary Information VII). It took an eight-GPU system less than five hours to obtain all results, even though some materials contain over 400 atoms per unit cell. Such efficiency enables the material design, searching, and optimization in a much larger design space, including alloys, interfaces, and even amorphous solids, with superior engineered phonon properties for thermal storage, energy conversion and harvesting, and superconductivity. In parallel, by taking advantage of the flexibility endowed by virtual nodes, other properties that are challenging to predict for a conventional GNN can be predicted similarly, such as electronic band structures and tight-binding and $k \cdot p$ effective Hamiltonian with a variable number of bands, optical properties like flexible optical absorption peaks as in the Lorentz oscillator model, and magnetic properties such as the number of propagation vectors. 

\begin{table}[ht!]
\begin{center}
\caption{Comparison of how the virtual nodes contribute to phonon prediction in terms of physics and computational costs. Here $m$, $N_{train}$, $t$, $n$ indicate the number of atoms per unit cell, the average of that in training data, the number of the unit cell counts, and an arbitrary, not large number respectively.}\label{table_phonon_methods}%
\begin{tabular}{c|cccc}
 & VVN & MVN & k-MVN \\
\hline\hline
\\
Force constants & - & - &  Reflected in VDM    \\
\\
Dynamical matrices & - & VDM & VDM    \\
\\
Phonon data & Virtual nodes & Eigenvalues & Eigenvalues   \\
\\
\hline
\\
Run time & $O(m^{2})$ & $O(m^{2.37})$ & $O(t\times m^{2.37})$   \\
\\
Storage & $O(m)$ & $O(m^{2})$ & $O(t\times m^{2})$  \\
\\
Generalization to larger systems & False & True & True  \\
\\
\hline
\end{tabular}
\end{center}
\end{table}

\nocitemain{*}
\bibliographystylemain{abbrv}
\bibliographymain{ms}

\section*{Methods}
\subsection*{Phonon data preparation}
We trained all of our models against an \textit{ab initio} DFPT computational database for phonon dispersion in harmonic model\cite{petretto2018high}. The data set contains material structures (the same as the primitive structure obtained from the Material Project\cite{jain2013commentary}), second-order derivatives of energies with respect to atomic perturbations for regular points inside the irreducible zone, and phonon dispersion along highly symmetric paths of 1,521 crystalline inorganic materials. These materials have 2 to 40 atoms per unit cell, with an average of 7.38. For this work, we only used the highly symmetric path phonon dispersion as our training data. The dispersion is between wave vectors $\vec{k}$ in the fractional reciprocal unit and response spectra in cm$^{-1}$. All models randomly split the data into 90\% training (1,365 materials), and 10\% testing (156 materials) sets. Furthermore, we trained our models with a 5-fold cross-validation scheme.

We also got phonon dispersion of complex (more number of atoms per unit cell) materials from Atsushi Togo's phonon database\cite{togo2015kyoto}. We used $seekpath$ \cite{hinuma2017band, togo2018texttt} module to get the highly symmetric path of each material. Then, we fed it alongside $POSCAR$, $FORCE\_SET$, and $phonopy.config$ files from the database to $Phonopy$\cite{togo2015first}'s python command to calculate the phonon dispersion along such path. To quality control the data, we selected materials whose lowest $\Gamma$-phonon band is higher than $-0.07$ cm$^{-1}$. We also filtered the material to get only the ones with more than 40 atoms per unit cell. Finally, we randomly selected 156 (the same as the number of data in the testing set for ease of comparison) out of 505 filtered materials. We used them as our complex material data set.  

\subsection*{Computation environments}\label{Methods}
We coded the models in Python 3.9.13 and trained them on our GPU cluster with CUDA version 10.2. To facilitate the model implementation, and training, we used some important python modules: $Pymatgen$\cite{ong2013python} and $ase$\cite{mortensen2017atomic} for handling material structure files ($.CIF$), $PyTorch$\cite{pytorch2018pytorch} for managing model training framework, $e3nn$\cite{geiger2022e3nn} for implementing our neural network models in the form that is equivariant for Euclidean group.

\subsection*{Virtual node graph neural network (VGNN)}
We have developed a scheme for a graph neural network (GNN) for it to be able to have variable output dimensions depending on the input size. For ease of understanding, we will explain the method with our work on phonon prediction. 

Considering a material with $m$ atoms per unit cell, we add $n$ additional virtual atoms. We can adjust the number $n$ depending on the model architecture. Using both real and virtual atoms, we convert the crystal structures into periodic graphs with $m$ real nodes for the actual atoms and $n$ virtual nodes for the added virtual atoms. Then, we connect nodes with edges indicating the message-passing process. To preserve the structural information of the materials and limit the computational cost, we apply the following rules for connections. First, if the distance between the two real nodes is within a specified cutoff radius $r_\text{max}$, the real nodes are connected through bi-directed edges. We also set up an edge between a real node and a virtual node according to the model description, but this edge is directed from real to virtual nodes. Lastly, we embed the information of radial distance vector, e.g., $\vec{r}_{ab}$ from atom $b$ to $a$, in the form of radial basis functions and spherical harmonics on the corresponding edge as edge attributes, which represent the distance and the direction of $\vec{r}_{ab}$ respectively. 

Since each node represents an atom in the unit cell, we embedded the atomic numbers $A$ information as node attributes $\mathscr{A}$ by passing one-hot representation vectors of length 118 through an embedding layer. As for the model's input, we embedded the atomic masses $Z$ information as input node features $\mathscr{Z}$ by passing the product of atomic mass and one-hot representation of atomic number through an embedding layer.

The constructed graph is then passed through the model message passing that operates on the features with multiple convolutions and gated activation layers\cite{miller2020relevance}. After the final layer, which consists of only a convolution (no gated activation), each of the $n$ virtual node features is collected, and passed through the post-processing block, which output the $3m$ predicted phonon branches. The post-processing block is different and will be explained in detail in the subsequent section of each model.

The model is optimized by minimizing the mean squared error (MSE) loss function between the phonon of the training data set and the one predicted by the model after normalizing them by the maximum phonon frequency of each material. The full network structure is provided in the supplementary Information.

\subsection*{Vector virtual node method (VVN)}
VVN is a VGNN we designed for learning to predict $\Gamma$-phonon spectra from material structures. Since, for a material with $m$ atoms per unit cell, there are $3m$ phonon bands, one sensible choice of adding virtual atoms is to add $3m$ virtual atoms each outputs the prediction of one of the bands. Hence, when there are $m$ atoms in the unit cell of crystalline material, we assign the position $\vec{r}_{V_i}$ of the virtual nodes $V_{i}, i\in[1, 3m]$ following equation (\ref{eq_vvn}). We can set the atomic species of the virtual node as anything, and we use Fe after optimization. 
\begin{equation}
\vec{r}_{V_i} = \frac{i-1}{3m} (\vec{a} + \vec{b}  + \vec{c}).
\label{eq_vvn}
\end{equation}
Here $\vec{a}$, $\vec{b}$, $\vec{c}$ indicates the unit cell vector of the material. In other words, $3m$ virtual atoms are placed along the diagonal line from $(0, 0, 0)$ to $\vec{a} + \vec{b}  + \vec{c}$ with equal spacing. By keeping the distances between the virtual nodes in the real space, it is possible to give position dependencies to the feature updating process. In that sense, equation (\ref{eq_vvn}) can consistently keep virtual nodes away from each other and enables us to use the virtual $3m$ virtual nodes as the output nodes of the network. To get information from the whole structure, each of the $3m$ virtual nodes is connected to all of the real nodes via directed edges from real to virtual nodes. After each convolution layer, the virtual node features are passed to a linear layer, converted to a scalar output, and sorted based on their magnitudes. The outputted $3m$ scalars represent the predicted $\Gamma$-phonon.

\subsection*{Matrix virtual node method (MVN)}
MVN is a VGNN we designed with the influence of the dynamic matrix representation of a periodic harmonic system for learning to predict $\Gamma$-phonon spectra from material structures. Given the momentum vector $\vec{k}$, the dynamical matrix element $\Tilde{D}_{ij}(\vec{k})$, which is a three-by-three matrix representing 3D harmonic interaction between atom $R_i$ and $R_j$, can be written as the Fourier transform of the force constant matrix $\Phi^{\alpha\beta}_{ij}$ following equation (\ref{eq_dynamical}). Here, $Z_{R_i}$ is $R_i$ atom's atomic mass, and $\vec{T}_\alpha$ is the $\alpha^\text{th}$ unit cell position. Note that, for each $k$-vector, the system has $3m$ degrees of freedom and frequencies where $m$ is the number of atoms per unit cell. We can get the phonon dispersion relations $\omega(\vec{k})$ by solving eigenvalues $\omega^2(\vec{k})$ of $\Tilde{D}(\vec{k})$, which is a matrix with shape $(3m, 3m)$ that composed of $m^2$ blocks of $\Tilde{D}_{ij}(\vec{k})$ for $i, j\in [1, m]$,
\begin{equation}
\Tilde{D}_{ij}(\vec{k}) = \sum_{\alpha,\beta}{\frac{\Phi^{\alpha\beta}_{ij}}{\sqrt{Z_{R_i}Z_{R_j}}}e^{i\vec{k}\cdot(\vec{T}_\alpha-\vec{T}_\beta)}}.   \\
\label{eq_dynamical}
\end{equation}
In the MVN method, we generate a matrix that could work like a dynamical matrix as is written in equation (\ref{eq_dynamical}). Here, we focus on the prediction of $\Gamma$-phonon, i.e. $\vec{k}=\vec{0}$. So, the contributions of the same atom pair, e.g., $R_i$, and $R_j$ from every unit cell separation $\vec{T}_\alpha-\vec{T}_\beta$ are summed without the $\vec{k}$-dependent exponential phase factor. Hence, the model needs to predict a matrix with shape $(3m, 3m)$ representing such summation. In order to do that, while preserving the relation of each matrix element, we generate $m$ virtual crystals $C_{j}, j\in [1, m]$ each of which has $m$ virtual nodes $V_{ij}, i\in [1, m]$ of the same atomic species and at the same positions as the real atoms $R_{i}, i\in[1, m]$. Here, a virtual node $V_{ij}$ represents the interaction term from a real node $R_{j}$ to another real node $R_{i}$ by adding a directed edge from $R_{j}$ to $V_{ij}$ whenever there is an edge connecting $R_{j}$ to $R_{i}$. After each convolution layer, the virtual node features are passed to a linear layer and converted to complex-valued output vectors with length 9. For each output feature, we reshape the output features into three-by-three matrices and arrange them such that $V_{ij}$'s matrix is the $(i, j)$ block of $\Tilde{D}$ supermatrix with shape ($3m$, $3m$). Finally, we solve $\Tilde{D}$ for its $3m$ eigenvalues, which work as the $\Gamma$-phonon prediction.

\subsection*{Momentum-dependent matrix virtual node method ($k$-MVN)}
$k$-MVN is a generalization of MVN model with non-zero $\vec{k}$. Unlike the MVN case, the $k$-MVN model needs to predict matrices representing interactions between atoms from a unit cell, e.g., $\vec{T}_\beta$, to the different unit cell, e.g., $\vec{T}_\alpha$. Since the phase factor depends only on the difference in unit cell positions, we can redefine $\vec{T}$ to be such a difference and simplify equation (\ref{eq_dynamical}) into
\begin{equation}
\Tilde{D}_{ij}(\vec{k}) = \sum_{\vec{T}}{\frac{\Phi^{\vec{T}}_{ij}}{\sqrt{Z_{R_i}Z_{R_j}}}e^{i\vec{k}\cdot\vec{T}}}\coloneqq \sum_{\vec{T}}{D_{ij}^{\vec{T}}e^{i\vec{k}\cdot\vec{T}}}.
\label{eq_dynamical_sim}
\end{equation}
\\
With this simplification, for each $\vec{T}$, we generate $m$ virtual crystal $C_{j\in[1, m]}^{\vec{T}}$ the same way as in MVN. However, in this case, $V_{ij}^{\vec{T}}$ represents the interaction term from a real node $R_{j}$ to another real node $R_{i}$ that is in the unit cell with unit cell position $\vec{T}$ with respect to $R_{j}$'s. In other words, we add a directed edge from $R_{j}$ to $V_{ij}^{\vec{T}}$ whenever there is an edge connecting $R_{j}$ to $R_{i}$ and that edge represent $\vec{r}_i-\vec{r}_j = \vec{r}'_i+\vec{T}-\vec{r}'_j$. Here, $\vec{r}'$ is the atomic position relative to its unit cell. Since GNN only considers edges with interatomic distance less than $r_\text{max}$, the model can generate, with this scheme, a non-zero matrix for only a finite number of $\vec{T}$ that satisfy
\begin{equation}
\min_{i, j\in [1, m]}|\vec{r}'_i+\vec{T}-\vec{r}'_j|\leq r_\text{max}.
\label{T_criterion}
\end{equation}
Hence, before the virtual crystal generations, the model also iterates through atom pairs to find all viable $\vec{T}$. 

Similar to the MVN model, we convert virtual node features into three-by-three matrices and merge them into a matrix with shape $(3m, 3m)$ representing $\Tilde{D}^{\vec{T}}$ for each $\vec{T}$. Finally, we weight sum these matrices with their phase factor to get $\Tilde{D}$ and solve for its $3m$ eigenvalues as phonon spectrum at wave vector $\vec{k}$.\\

\noindent\textbf{Acknowledgements}\\
RO and AC contribute equally to this work. RO, AC, AB, and ML thank M Geiger, S Fang, T Smidt, and K Persson for helpful discussions, and acknowledge the support from the U.S. Department of Energy (DOE),  Office of Science (SC), Basic Energy Sciences (BES), Award No. DE-SC0021940, and National Science Foundation (NSF) Designing Materials to Revolutionize and Engineer our Future (DMREF) Program with Award No. DMR-2118448. BL acknowledges the support of NSF DMREF with Award No. DMR-2118523. TN, ND, and ML are partially supported by DOE BES Award No. DE-SC0020148. TN acknowledges support from Mathworks Fellowship and Sow-Hsin Chen Fellowship. ML acknowledges the support from the Class of 1947 Career Development Chair and discussions with S. Yip. 
\\

\noindent\textbf{Competing interests}\\
The authors declare no competing interests \\

\noindent\textbf{Data Availability Statement}\\
The data that support the findings of this study are openly available in GitHub at \url{https://github.com/RyotaroOKabe/phonon_prediction}. The $\Gamma$-phonon database generated with the MVN method is available at \url{https://osf.io/k5utb/} \\


\begin{thebibliography}{10}

\bibitem{altintas2021machine}
C.~Altintas, O.~F. Altundal, S.~Keskin, and R.~Yildirim.
\newblock Machine learning meets with metal organic frameworks for gas storage
  and separation.
\newblock {\em Journal of Chemical Information and Modeling}, 61(5):2131--2146,
  2021.

\bibitem{baroni:phonon:2001}
S.~Baroni, S.~Gironcoli, and A.~Corso.
\newblock Phonons and related crystal properties from density-functional
  perturbation theory.
\newblock {\em Rev. Mod. Phys.}, 73:515, 2001.

\bibitem{cheng2011atomic}
Y.~Cheng and E.~Ma.
\newblock Atomic-level structure and structure--property relationship in
  metallic glasses.
\newblock {\em Progress in materials science}, 56(4):379--473, 2011.

\bibitem{delaire2009phonon}
O.~Delaire, A.~F. May, M.~A. McGuire, W.~D. Porter, M.~S. Lucas, M.~B. Stone,
  D.~L. Abernathy, V.~Ravi, S.~Firdosy, and G.~Snyder.
\newblock Phonon density of states and heat capacity of la 3- x te 4.
\newblock {\em Physical Review B}, 80(18):184302, 2009.

\bibitem{dresselhaus2007new}
M.~S. Dresselhaus, G.~Chen, M.~Y. Tang, R.~Yang, H.~Lee, D.~Wang, Z.~Ren, J.-P.
  Fleurial, and P.~Gogna.
\newblock New directions for low-dimensional thermoelectric materials.
\newblock {\em Advanced materials}, 19(8):1043--1053, 2007.

\bibitem{dunn2020benchmarking}
A.~Dunn, Q.~Wang, A.~Ganose, D.~Dopp, and A.~Jain.
\newblock Benchmarking materials property prediction methods: the matbench test
  set and automatminer reference algorithm.
\newblock {\em npj Computational Materials}, 6(1):1--10, 2020.

\bibitem{fung2021benchmarking}
V.~Fung, J.~Zhang, E.~Juarez, and B.~G. Sumpter.
\newblock Benchmarking graph neural networks for materials chemistry.
\newblock {\em npj Computational Materials}, 7(1):1--8, 2021.

\bibitem{geiger2022e3nn}
M.~Geiger and T.~Smidt.
\newblock e3nn: Euclidean neural networks.
\newblock {\em arXiv preprint arXiv:2207.09453}, 2022.

\bibitem{guo2021artificial}
K.~Guo, Z.~Yang, C.-H. Yu, and M.~J. Buehler.
\newblock Artificial intelligence and machine learning in design of mechanical
  materials.
\newblock {\em Materials Horizons}, 8(4):1153--1172, 2021.

\bibitem{hart2021machine}
G.~L. Hart, T.~Mueller, C.~Toher, and S.~Curtarolo.
\newblock Machine learning for alloys.
\newblock {\em Nature Reviews Materials}, 6(8):730--755, 2021.

\bibitem{hinuma2017band}
Y.~Hinuma, G.~Pizzi, Y.~Kumagai, F.~Oba, and I.~Tanaka.
\newblock Band structure diagram paths based on crystallography.
\newblock {\em Computational Materials Science}, 128:140--184, 2017.

\bibitem{jain2013commentary}
A.~Jain, S.~P. Ong, G.~Hautier, W.~Chen, W.~D. Richards, S.~Dacek, S.~Cholia,
  D.~Gunter, D.~Skinner, G.~Ceder, et~al.
\newblock Commentary: The materials project: A materials genome approach to
  accelerating materials innovation.
\newblock {\em APL materials}, 1(1):011002, 2013.

\bibitem{jancar2010current}
J.~Jancar, J.~Douglas, F.~W. Starr, S.~Kumar, P.~Cassagnau, A.~Lesser, S.~S.
  Sternstein, and M.~Buehler.
\newblock Current issues in research on structure--property relationships in
  polymer nanocomposites.
\newblock {\em Polymer}, 51(15):3321--3343, 2010.

\bibitem{kong2011phonon}
L.~T. Kong.
\newblock Phonon dispersion measured directly from molecular dynamics
  simulations.
\newblock {\em Computer Physics Communications}, 182(10):2201--2207, 2011.

\bibitem{kumar2020topological}
N.~Kumar, S.~N. Guin, K.~Manna, C.~Shekhar, and C.~Felser.
\newblock Topological quantum materials from the viewpoint of chemistry.
\newblock {\em Chemical Reviews}, 121(5):2780--2815, 2020.

\bibitem{le2012quantitative}
T.~Le, V.~C. Epa, F.~R. Burden, and D.~A. Winkler.
\newblock Quantitative structure--property relationship modeling of diverse
  materials properties.
\newblock {\em Chemical reviews}, 112(5):2889--2919, 2012.

\bibitem{liu2018antiferroelectrics}
Z.~Liu, T.~Lu, J.~Ye, G.~Wang, X.~Dong, R.~Withers, and Y.~Liu.
\newblock Antiferroelectrics for energy storage applications: a review.
\newblock {\em Advanced Materials Technologies}, 3(9):1800111, 2018.

\bibitem{miller2020relevance}
B.~K. Miller, M.~Geiger, T.~E. Smidt, and F.~No{\'e}.
\newblock Relevance of rotationally equivariant convolutions for predicting
  molecular properties.
\newblock {\em arXiv preprint arXiv:2008.08461}, 2020.

\bibitem{mishra2009metal}
A.~Mishra, M.~K. Fischer, and P.~B{\"a}uerle.
\newblock Metal-free organic dyes for dye-sensitized solar cells: From
  structure: Property relationships to design rules.
\newblock {\em Angewandte Chemie International Edition}, 48(14):2474--2499,
  2009.

\bibitem{mortensen2017atomic}
J.~Mortensen, J.~Blomqvist, I.~Castelli, R.~Christensen, M.~Du{\l}ak, J.~Friis,
  M.~Groves, B.~Hammer, C.~Hargus, E.~Hermes, et~al.
\newblock The atomic simulation environment-a python library for working with
  atoms.
\newblock {\em Journal of physics. Condensed Matter: an Institute of Physics
  Journal}, 29(27):273002--273002, 2017.

\bibitem{oganov2006crystal}
A.~R. Oganov and C.~W. Glass.
\newblock Crystal structure prediction using ab initio evolutionary techniques:
  Principles and applications.
\newblock {\em The Journal of chemical physics}, 124(24):244704, 2006.

\bibitem{oganov2019structure}
A.~R. Oganov, C.~J. Pickard, Q.~Zhu, and R.~J. Needs.
\newblock Structure prediction drives materials discovery.
\newblock {\em Nature Reviews Materials}, 4(5):331--348, 2019.

\bibitem{ong2013python}
S.~P. Ong, W.~D. Richards, A.~Jain, G.~Hautier, M.~Kocher, S.~Cholia,
  D.~Gunter, V.~L. Chevrier, K.~A. Persson, and G.~Ceder.
\newblock Python materials genomics (pymatgen): A robust, open-source python
  library for materials analysis.
\newblock {\em Computational Materials Science}, 68:314--319, 2013.

\bibitem{peng2022human}
J.~Peng, D.~Schwalbe-Koda, K.~Akkiraju, T.~Xie, L.~Giordano, Y.~Yu, C.~J. Eom,
  J.~R. Lunger, D.~J. Zheng, R.~R. Rao, et~al.
\newblock Human-and machine-centred designs of molecules and materials for
  sustainability and decarbonization.
\newblock {\em Nature Reviews Materials}, pages 1--19, 2022.

\bibitem{petretto2018high}
G.~Petretto, S.~Dwaraknath, H.~PC~Miranda, D.~Winston, M.~Giantomassi, M.~J.
  Van~Setten, X.~Gonze, K.~A. Persson, G.~Hautier, and G.-M. Rignanese.
\newblock High-throughput density-functional perturbation theory phonons for
  inorganic materials.
\newblock {\em Scientific data}, 5(1):1--12, 2018.

\bibitem{pytorch2018pytorch}
A.~D.~I. Pytorch.
\newblock Pytorch, 2018.

\bibitem{schwalbe2021priori}
D.~Schwalbe-Koda, S.~Kwon, C.~Paris, E.~Bello-Jurado, Z.~Jensen, E.~Olivetti,
  T.~Willhammar, A.~Corma, Y.~Rom{\'a}n-Leshkov, M.~Moliner, et~al.
\newblock A priori control of zeolite phase competition and intergrowth with
  high-throughput simulations.
\newblock {\em Science}, 374(6565):308--315, 2021.

\bibitem{sham1969electronic}
L.~Sham.
\newblock Electronic contribution to lattice dynamics in insulating crystals.
\newblock {\em Physical Review}, 188(3):1431, 1969.

\bibitem{stanev2021artificial}
V.~Stanev, K.~Choudhary, A.~G. Kusne, J.~Paglione, and I.~Takeuchi.
\newblock Artificial intelligence for search and discovery of quantum
  materials.
\newblock {\em Communications Materials}, 2(1):1--11, 2021.

\bibitem{Thomas2018TFN}
N.~{Thomas}, T.~{Smidt}, S.~{Kearnes}, L.~{Yang}, L.~{Li}, K.~{Kohlhoff}, and
  P.~{Riley}.
\newblock {Tensor field networks: Rotation- and translation-equivariant neural
  networks for 3D point clouds}.
\newblock {\em arXiv e-prints}, page arXiv:1802.08219, Feb. 2018.

\bibitem{togo2015kyoto}
A.~Togo.
\newblock Phonon database at kyoto university.
\newblock {\em http://phonondb.mtl.kyoto-u.ac.jp/}, 2015.

\bibitem{togo2015first}
A.~Togo and I.~Tanaka.
\newblock First principles phonon calculations in materials science.
\newblock {\em Scripta Materialia}, 108:1--5, 2015.

\bibitem{togo2018texttt}
A.~Togo and I.~Tanaka.
\newblock $spglib$ : a software library for crystal symmetry search.
\newblock {\em arXiv preprint arXiv:1808.01590}, 2018.

\bibitem{wagih2020learning}
M.~Wagih, P.~M. Larsen, and C.~A. Schuh.
\newblock Learning grain boundary segregation energy spectra in polycrystals.
\newblock {\em Nature communications}, 11(1):1--9, 2020.

\bibitem{xie2018crystal}
T.~Xie and J.~C. Grossman.
\newblock Crystal graph convolutional neural networks for an accurate and
  interpretable prediction of material properties.
\newblock {\em Physical review letters}, 120(14):145301, 2018.

\bibitem{yao2022high}
Y.~Yao, Q.~Dong, A.~Brozena, J.~Luo, J.~Miao, M.~Chi, C.~Wang, I.~G.
  Kevrekidis, Z.~J. Ren, J.~Greeley, et~al.
\newblock High-entropy nanoparticles: Synthesis-structure-property
  relationships and data-driven discovery.
\newblock {\em Science}, 376(6589):eabn3103, 2022.

\bibitem{zheng2021metal}
W.~Zheng and L.~Y.~S. Lee.
\newblock Metal--organic frameworks for electrocatalysis: catalyst or
  precatalyst?
\newblock {\em ACS Energy Letters}, 6(8):2838--2843, 2021.

\bibitem{zhu2021charting}
T.~Zhu, R.~He, S.~Gong, T.~Xie, P.~Gorai, K.~Nielsch, and J.~C. Grossman.
\newblock Charting lattice thermal conductivity for inorganic crystals and
  discovering rare earth chalcogenides for thermoelectrics.
\newblock {\em Energy \& Environmental Science}, 14(6):3559--3566, 2021.

\end{thebibliography}


\begin{thebibliography}{1}

\bibitem{bellaiche2000virtual}
L.~Bellaiche and D.~Vanderbilt.
\newblock Virtual crystal approximation revisited: Application to dielectric
  and piezoelectric properties of perovskites.
\newblock {\em Physical Review B}, 61(12):7877, 2000.

\bibitem{chen2021dos}
Z.~Chen, N.~Andrejevic, T.~Smidt, Z.~Ding, Q.~Xu, Y.-T. Chi, Q.~T. Nguyen,
  A.~Alatas, J.~Kong, and M.~Li.
\newblock Direct prediction of phonon density of states with euclidean neural
  networks.
\newblock {\em Advanced Science}, 8(12):2004214, 2021.

\bibitem{geiger2022e3nn}
M.~Geiger and T.~Smidt.
\newblock e3nn: Euclidean neural networks.
\newblock {\em arXiv preprint arXiv:2207.09453}, 2022.

\bibitem{jain2013commentary}
A.~Jain, S.~P. Ong, G.~Hautier, W.~Chen, W.~D. Richards, S.~Dacek, S.~Cholia,
  D.~Gunter, D.~Skinner, G.~Ceder, et~al.
\newblock Commentary: The materials project: A materials genome approach to
  accelerating materials innovation.
\newblock {\em APL materials}, 1(1):011002, 2013.

\bibitem{kormann2017phonon}
F.~K{\"o}rmann, Y.~Ikeda, B.~Grabowski, and M.~H. Sluiter.
\newblock Phonon broadening in high entropy alloys.
\newblock {\em npj Computational materials}, 3(1):1--9, 2017.

\bibitem{Thanh2020Kohn}
T.~Nguyen, F.~Han, N.~Andrejevic, R.~Pablo-Pedro, A.~Apte, Y.~Tsurimaki,
  Z.~Ding, K.~Zhang, A.~Alatas, E.~E. Alp, S.~Chi, J.~Fernandez-Baca,
  M.~Matsuda, D.~A. Tennant, Y.~Zhao, Z.~Xu, J.~W. Lynn, S.~Huang, and M.~Li.
\newblock Topological singularity induced chiral kohn anomaly in a weyl
  semimetal.
\newblock {\em Phys. Rev. Lett.}, 124:236401, Jun 2020.

\bibitem{petretto2018high}
G.~Petretto, S.~Dwaraknath, H.~PC~Miranda, D.~Winston, M.~Giantomassi, M.~J.
  Van~Setten, X.~Gonze, K.~A. Persson, G.~Hautier, and G.-M. Rignanese.
\newblock High-throughput density-functional perturbation theory phonons for
  inorganic materials.
\newblock {\em Scientific data}, 5(1):1--12, 2018.

\bibitem{togo2015kyoto}
A.~Togo.
\newblock Phonon database at kyoto university.
\newblock {\em http://phonondb.mtl.kyoto-u.ac.jp/}, 2015.

\end{thebibliography}
\end{document}


\maketitle
\newcommand{\beginsupplement}{%
        \setcounter{table}{0}
        \renewcommand{\thetable}{S\arabic{table}}%
        \setcounter{figure}{0}
        \renewcommand{\thefigure}{S\arabic{figure}}%
     }
\beginsupplement
\tableofcontents
\section{DATA PREPARATION}

\begin{figure}[h!]
\includegraphics[width=\textwidth]{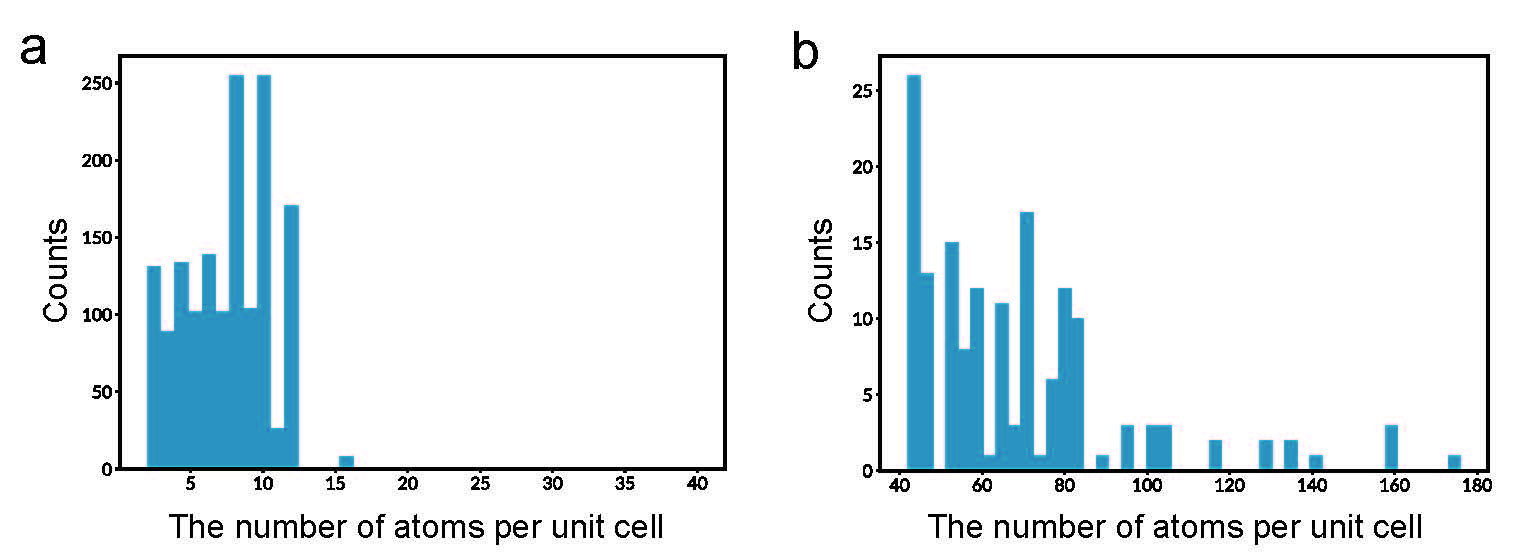}
\centering
\caption{\textbf{The distribution of atoms per unit cell.} 
The number of atoms per unit cell in \textbf{a}. Training and test data set from the Main Database \textbf{b}. Additional test data set of complex materials from the Togo Database. \textbf{a}. The total 1521 materials from the Main database contain 2 to 40 atoms per unit cell with an average of 7.4 atoms per unit cell. \textbf{b}. The randomly selected 156 complex materials with good ground-truth quality in the Togo Database with 42 to 174 atoms per unit cell (an average of 69.1 atoms per unit cell).
}
\label{hist_sites}
\end{figure}

\begin{figure}[h!]
\includegraphics[width=\textwidth]{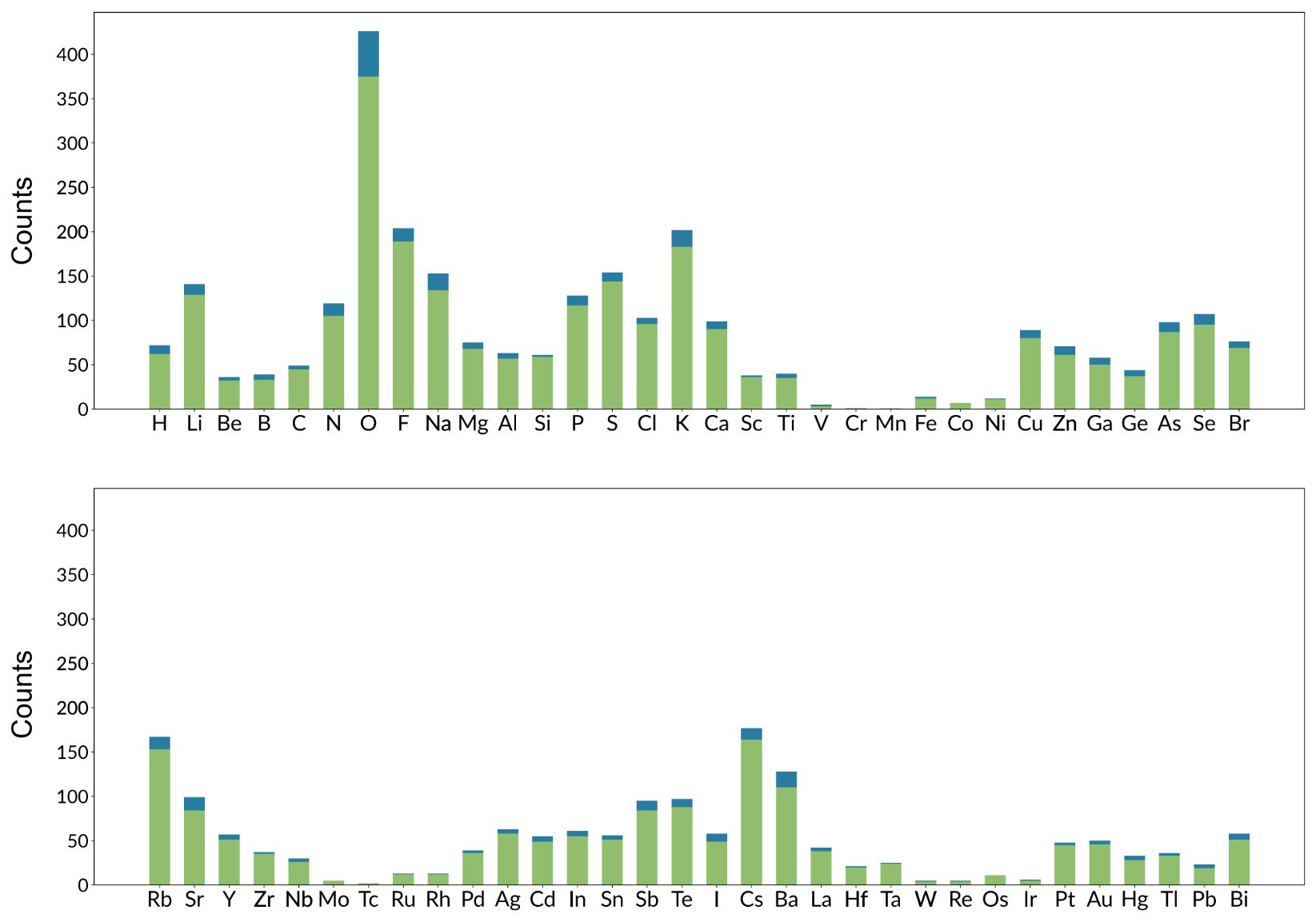}
\centering
\caption{\textbf{Training and testing data by elements in the Main Database.} 
The number of appearances by each chemical element in the training (green) and the testing (blue) data sets. 
}
\label{element_count_train}
\end{figure}

\begin{figure}[h!]
\includegraphics[width=\textwidth]{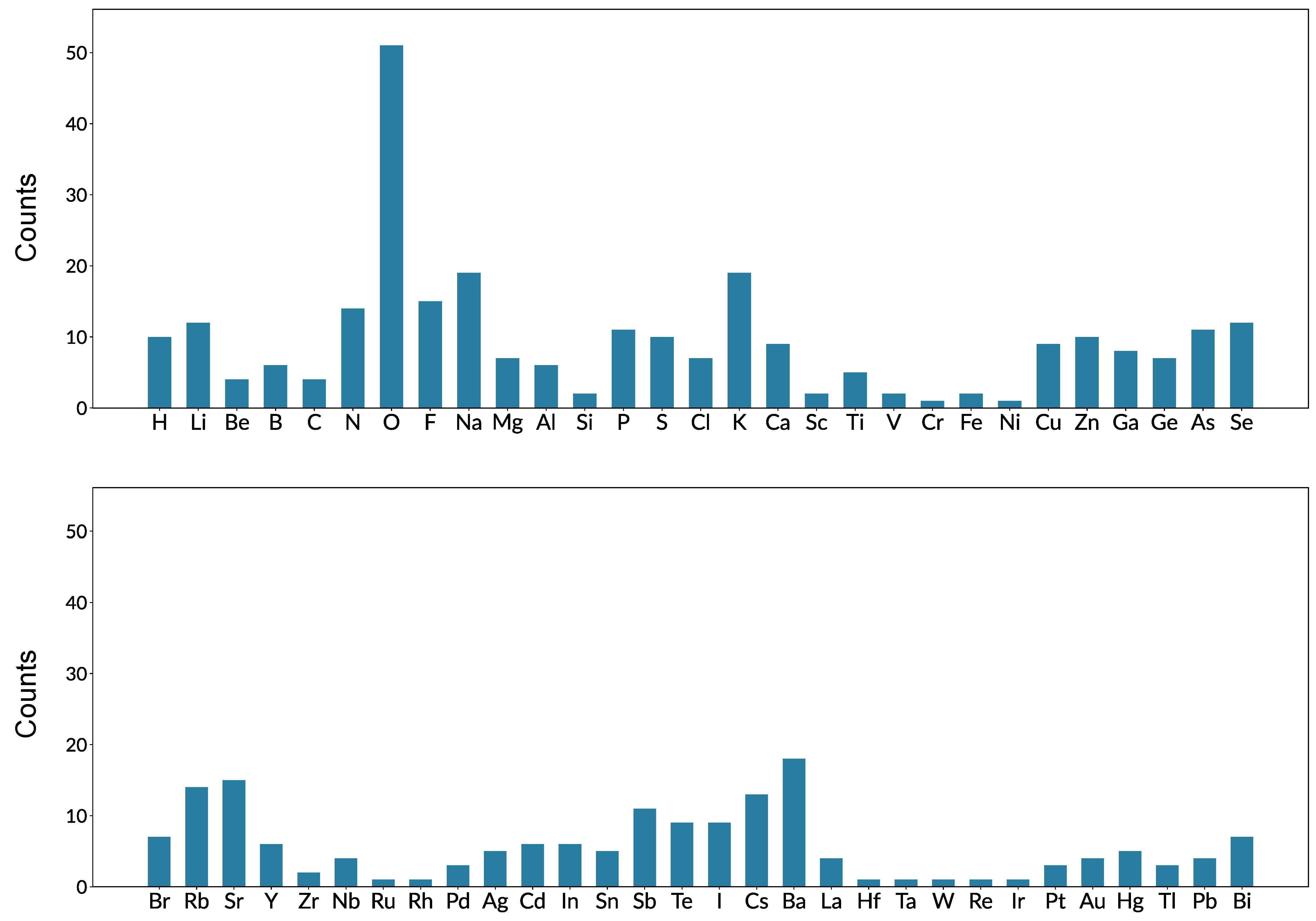}
\centering
\caption{\textbf{Additional testing data by elements in the Togo Database.} 
The number of appearances by each chemical element in testing data of complex materials is shown among the 156 randomly selected materials. 
}
\label{element_count_kyoto}
\end{figure}

The data set containing full phonon bands of 1521 semiconducting inorganic materials are calculated from density functional perturbation theory (DFPT) \cite{petretto2018high}, which is the main dataset for training (``Main Database" for short). The phonon energy values at $\Gamma$-points (for vector virtual nodes, VVN and matrix virtual nodes, MVN model), and other high symmetric points ($k$-MVN model) were extracted and randomly split for training (90\%) and testing set (10\%). High symmetric points are different for each material depending on its structure, and the reduced fractional coordinates are implemented for both real and reciprocal spaces for a primitive unit cell. A consistency check has been performed for all data, ensuring the intended $k$-points in the fractional unit match the desired $k$-points in the Brillouin zone. Another set of data for testing the model was taken from the Phonon database by Dr. Atsuhi Togo at Kyoto University (``Togo Database" for short)\cite{togo2015kyoto}, which contains the phonons of more complex materials, but meanwhile contain more phonons with imaginary phonon energies. We use the Main Database for training, given the more stringent convergence criteria. In addition, we randomly selected 156 materials in the Togo Database with the lowest $\Gamma$ phonons greater than -0.07 cm$^{-1}$, and have at least 40 atoms per unit cell for testing the model trained from the Main Database. The 156 materials from the Togo Database match the number as the testing set from the Main Database (10\% of the 1521 materials). Most elements appear in both the test set of the Main Database and the additional test in the Togo Database. The distribution of the number of atoms per unit cell ($m$-value in the main text) for the Main Database and the Togo Database are shown in Figure \ref{hist_sites}. The profiles of these data sets with respect to the elements are shown in Figures \ref{element_count_train} - \ref{element_count_kyoto}. 

\section{VIRTUAL NODE GRAPH NEURAL NETWORKS}
\subsection{Euclidean Neural Networks and Real Node Graph Convolutions}
Our approaches were developed within the framework of a symmetry-aware graph convolutional neural network, the Euclidean neural networks (E3NN)\cite{geiger2022e3nn}. Figure \ref{model-architecture} illustrates the overall architecture of the model. The model takes atomic mass and position of atoms in a unit cell as an input, the same as the previous report\cite{chen2021dos}. This information is converted to a periodic graph with nodes representing atoms and edges controlling the message passing between nodes. This input is then passed through a series of convolution layers separated by nonlinear layers, which introduce the complexity to the model. The convolution layer computes the tensor product of input features, and the convolution kernel is defined as:
\begin{equation}
f'_i = \frac{1}{\sqrt{z}}\sum_{j\in\partial(i)} f_j \otimes (h(||r_{ij} ||))Y(r_{ij}/|| {r_{ij}} ||)
\label{eq_convolution}
\end{equation}

\noindent where $f'_i$ is the output feature of atom $i$, and $\partial(i)$ is the set of neighbouring atoms which $r_{ij}\leq r_{max}$ where $r_{ij}$ is the relative position from atom j to i. We sum the tensor product of the features of atom $j$ and the convolution kernel consisting of the learned radial function $(h(||r_{ij} ||))$ and the spherical harmonics $Y(r_{ij}/|| {r_{ij}} ||)$. The normalizing factor $z$ is the coordination number, aka the number of neighboring atoms. 
After passing through the last convolution layer, the output features are processed through the screening virtual node and processing layer. The mean square error (MSE) loss is calculated and used for backpropagation to optimize the model. 

\subsection{Virtual Nodes Augmentation and Convolutions}
With the advantages of parameter sharing of graph convolution and symmetry awareness of Euclidean neural networks, the augmentation of the virtual node must satisfy the basic requirements for the model to still perform correct convolution that is equivariant for Euclidean group; Virtual nodes must be embedded with the same dimensions of node attributes, and input features as the real nodes and every edge connecting virtual nodes to any other nodes must be embedded with the same dimensions of edge attributes as the edges connecting real nodes.

In our work on phonon prediction, we decided to augment virtual nodes by adding virtual atoms that represent virtual nodes into the crystal structure and constructing the graph in the same way as when there are no virtual atoms. Because the graph construction treats virtual atoms and real atoms equally, the results must already satisfy all requirements. Although this augmentation method limits the freedom of virtual nodes to be constructed from some virtual atoms, there are still plenty of degrees of freedom for us to engineer the model construction: numbers, types, and positions of added virtual atoms. In fact, depending on the model, the virtual atoms do not need to be actual elements, e.g. virtual atoms with an atomic number of 4, but with an atomic mass of 0.

Another degree of freedom that the virtual node method allows is node connectivity. However, since we want the model to preserve the connection structure and message passing between real nodes, we decided to restrict the edges between real and virtual nodes to be directed from a real to a virtual node only, while the connections among real nodes are the same as previously described. Hence, with the restriction on the direction of message passing between real and virtual nodes, we can design our model with any remaining connection combinations.

To sum up, all of our models (VVN, MVN, and $k$-MVN) augment virtual nodes by adding a certain number of virtual atoms of certain types at certain locations in the original crystal cells. Then, they build a graph by treating virtual atoms as real atoms to keep the functionality of convolution and symmetry awareness. Finally, they include all edges connecting real nodes, include only some directed edges from real to virtual nodes, and include some edges connecting virtual nodes depending on the design of the model.

\subsection{Neural Network Architectures}
\begin{figure}[h!]
\includegraphics[height=0.9\textheight]{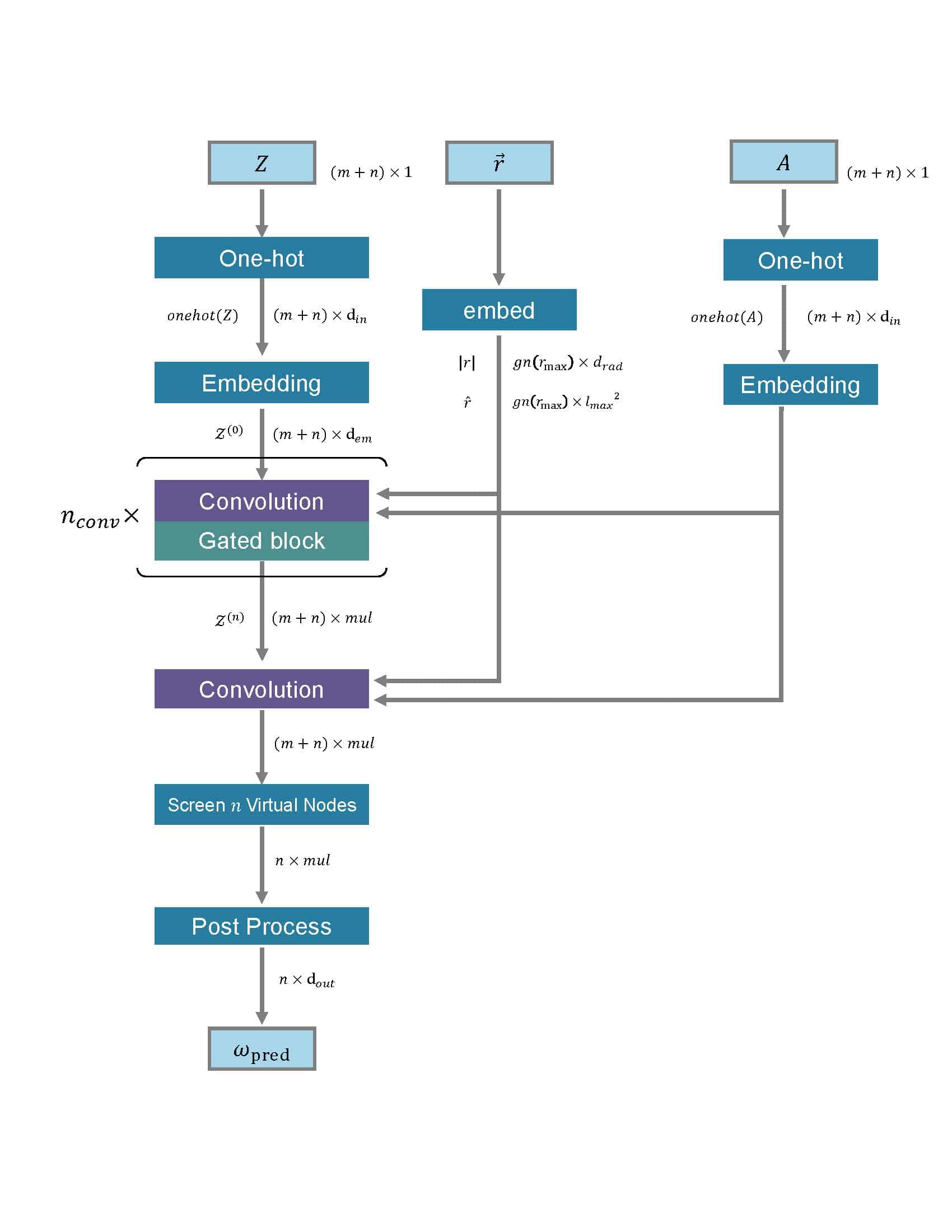}
\centering
\caption{\textbf{Overall architecture of the equivariant neural network and convolution layer.}
The model consists of n layers of convolution layer separated by non-linear layers, screen virtual node layer, and post-processing layer.
}
\label{model-architecture}
\end{figure}

The generic neural network architectures is shown in Figure \ref{model-architecture}.     \\ 

\newpage
\section{NEURAL NETWORK TRAINING AND OPTIMIZATION}
We optimize the values of the training parameters. The set of parameters that gives the best results for VVN, MVN, k-MVN are shown in Table \ref{hparams}.

\begin{table}[h!]
\begin{center}
\caption{The parameter setting of VVN}\label{hparams}
\begin{tabular}{cccc}
 Hyperparameter & VVN & MVN & $k$-MVN \\
\hline\hline
Maximum cutoff radius ($r_{max} [\mathring{A}]$)   & 4 & 4 & 4 \\
Multiplicity of irreducible representation ($mul$) & 16 & 4 & 4   \\
Number of pointwise convolution layer ($n_{conv}$) & 2  & 3 & 2    \\
Number of basis for radial filters ($n_{rad}$) & 10 & 10 & 10     \\
Maximum $l$ of spherical harmonics ($l_{max}$) & 2 & 2 & 2     \\
Length of embedding feature vector ($dim_{em}$) & 16 & 32 & 32   \\
Length of output feature vector ($dim_{out}$) & 1 & 18 & 18   \\
AdamW optimizer learning rate & $(5\cdot10^{-3})\times0.96^k$ & $(5\cdot10^{-3})\times0.96^k$ & $(5\cdot10^{-3})\times0.96^k$    \\
AdamW optimizer weight decay coefficient & 0.05 & 0.05 & 0.05    \\
\hline
\end{tabular}
\end{center}
\end{table}

\section{PERFORMANCE on $\Gamma$-PHONON PREDICTIONS}
\subsection{Additional $\Gamma$-Phonon Predictions}
\begin{figure}[H]
\includegraphics[height=0.9\textheight]{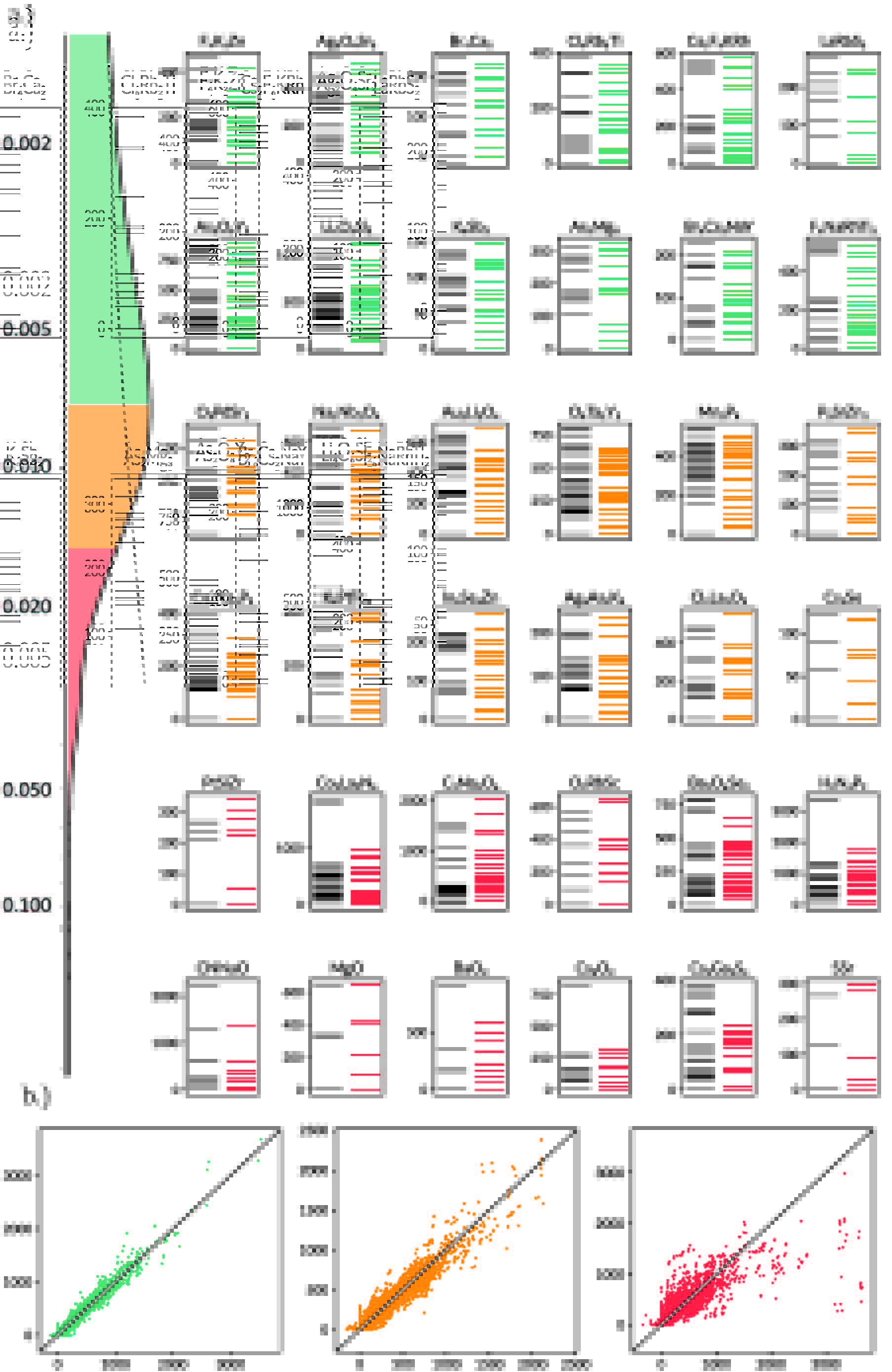}
\centering
\caption{\textbf{VVN model: Direct prediction results within the training set of the Main Database.} 
\textbf{a}. Black lines indicate the $\Gamma$-phonon from DFPT calculation (ground truth) in [$cm^{-1}$]. The colored (green, yellow, red) lines represent predicted $\Gamma$-phonon in 1st, 2nd, and 3rd tertiles, respectively. (Left) Loss distribution shows that it is heavily peaked in the 1st and 2nd tertiles with lower error.
\textbf{b}. The correlation plots (prediction Vs. ground truth) of all $\Gamma$-phonon within the training set in each tertile.  
}
\label{diag_train_pred}
\end{figure}

\begin{figure}[H]
\includegraphics[height=0.9\textheight]{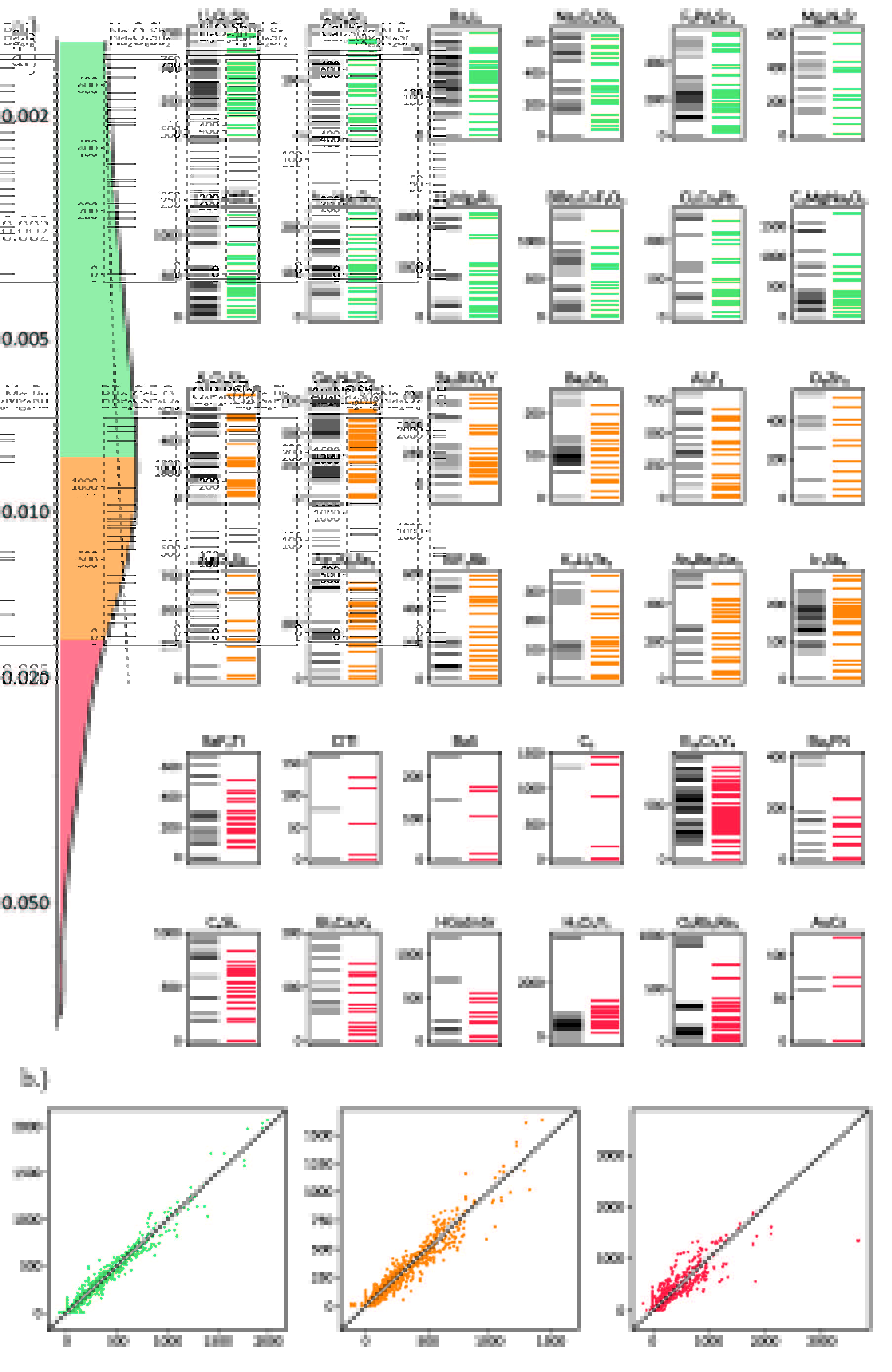}
\centering
\caption{\textbf{VVN model: Direct prediction results within the testing set of the Main Database.} 
\textbf{a}. The predictions of $\Gamma$-phonon comparing with DFPT calculations (black) in 1st, 2nd, and 3rd tertiles (green, yellow, and red) in [$cm^{-1}$]. The loss distribution on the left shows a peak in the 1st and 2nd tertiles.
\textbf{b}. The correlation plots (prediction Vs. ground truth) of all $\Gamma$-phonon within the test set in each tertile.  
}
\label{diag_test_pred}
\end{figure}

\begin{figure}[H]
\includegraphics[height=0.9\textheight]{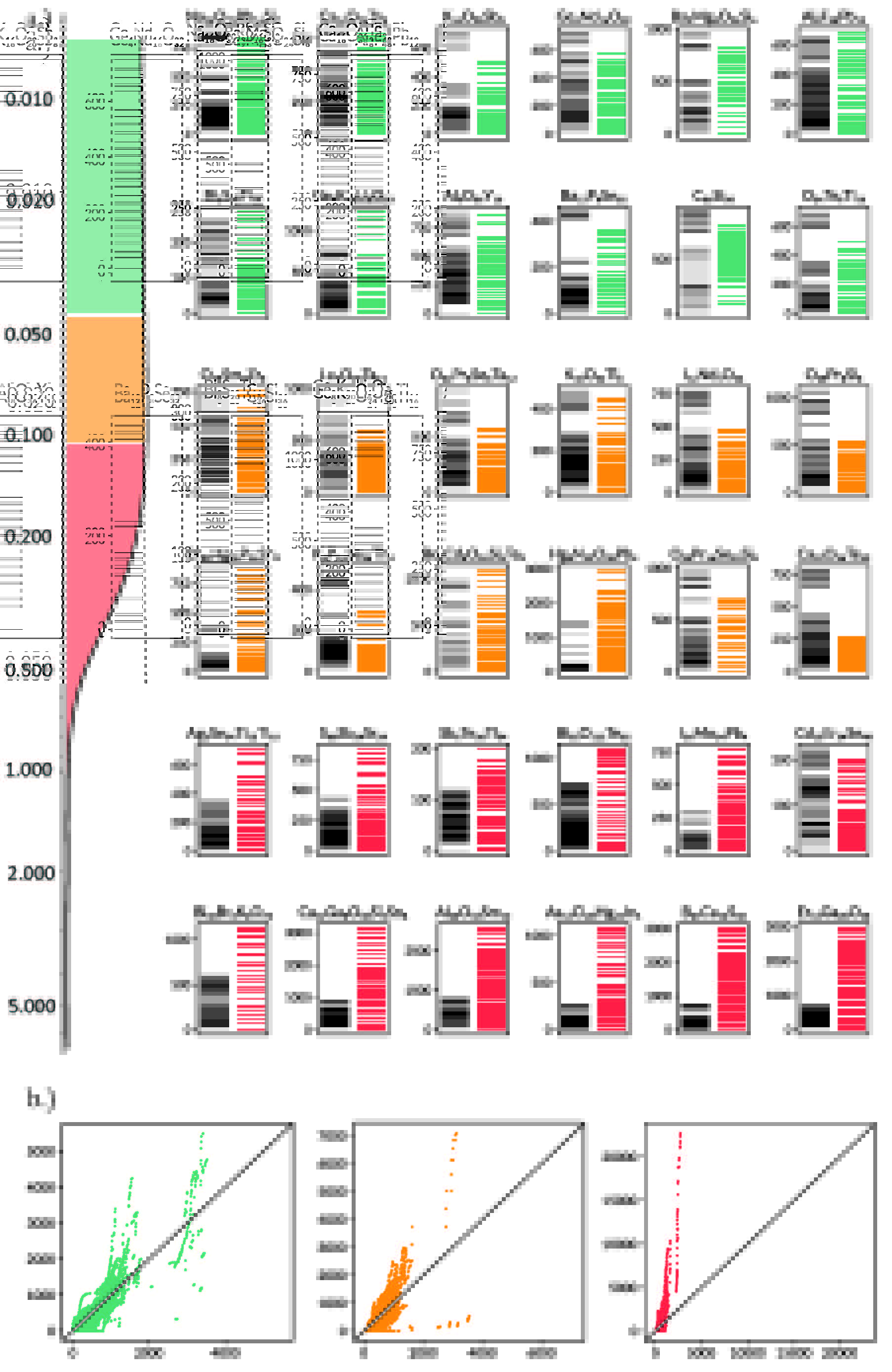}
\centering
\caption{\textbf{VNN model: Direct prediction results within test set of the Togo Database using VVN.} 
\textbf{a}. The predictions of $\Gamma$-phonon comparing with DFPT calculations (black) in 1st, 2nd, and 3rd tertiles (green, yellow, and red) in [$cm^{-1}$]. The loss distribution ranges broader in the higher values than those with training data (Figure \ref{diag_train_pred}) and testing data (Figure \ref{diag_test_pred}).
\textbf{b}. The correlation plots (prediction Vs. ground truth) of all $\Gamma$-phonon within the test set in each tertile.  
}
\label{diag_kyoto_pred}
\end{figure}

\begin{figure}[H]
\includegraphics[height=0.9\textheight]{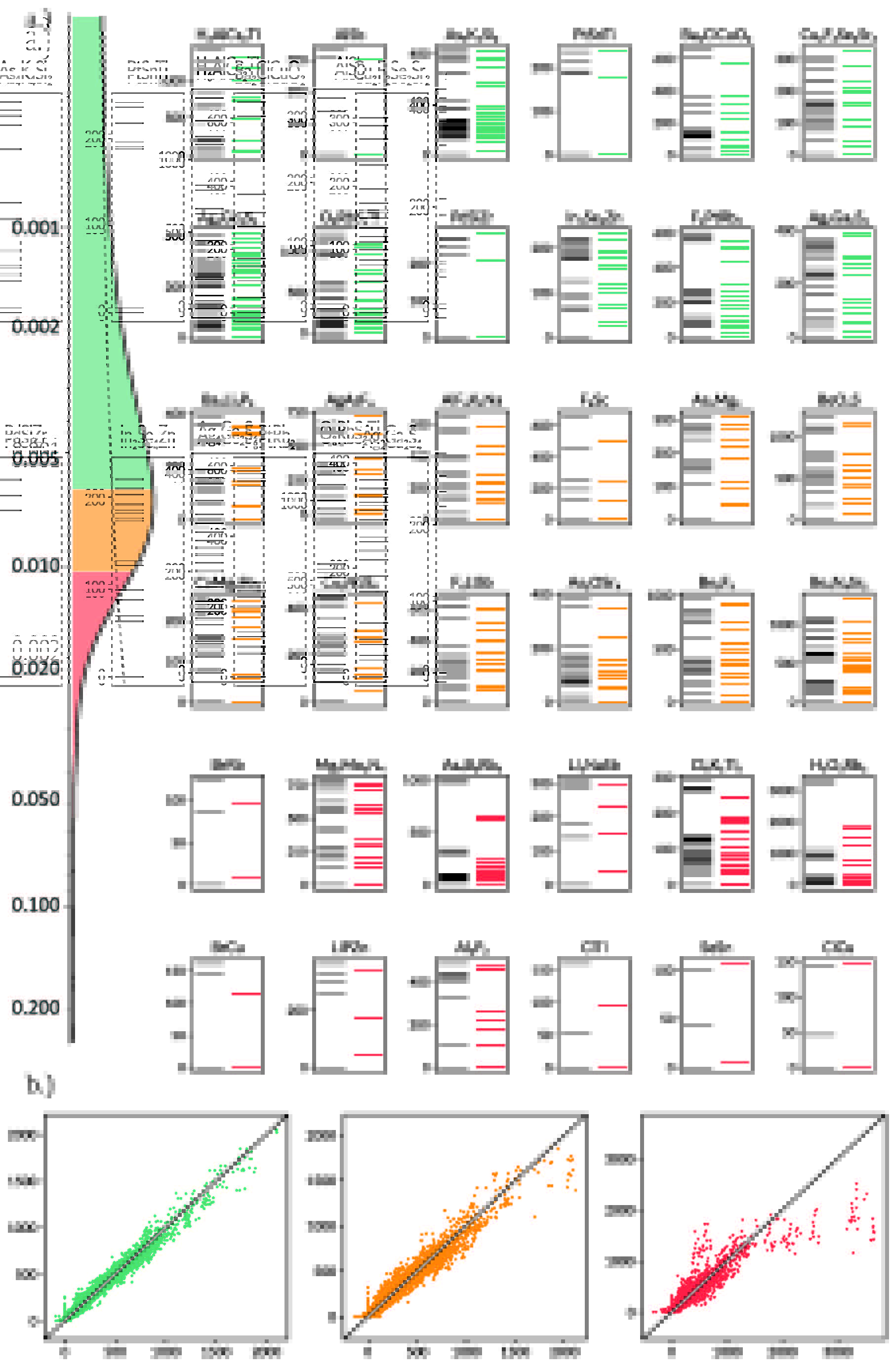}
\centering
\caption{\textbf{MVN model: Direct prediction results within train set of the Main Database using MVN} 
\textbf{a}. The predictions of $\Gamma$-phonon comparing with DFPT calculations (black) in 1st, 2nd and 3rd tertiles(green, yellow, and red) in [$cm^{-1}$]. The loss distribution on the left shows a peal in the 1st and 2nd tertiles.
\textbf{b}. The correlation plots (prediction Vs. ground truth) of all $\Gamma$-phonon within the test set in each tertile.  
}
\label{dynam_train_pred}
\end{figure}

\begin{figure}[H]
\includegraphics[height=0.9\textheight]{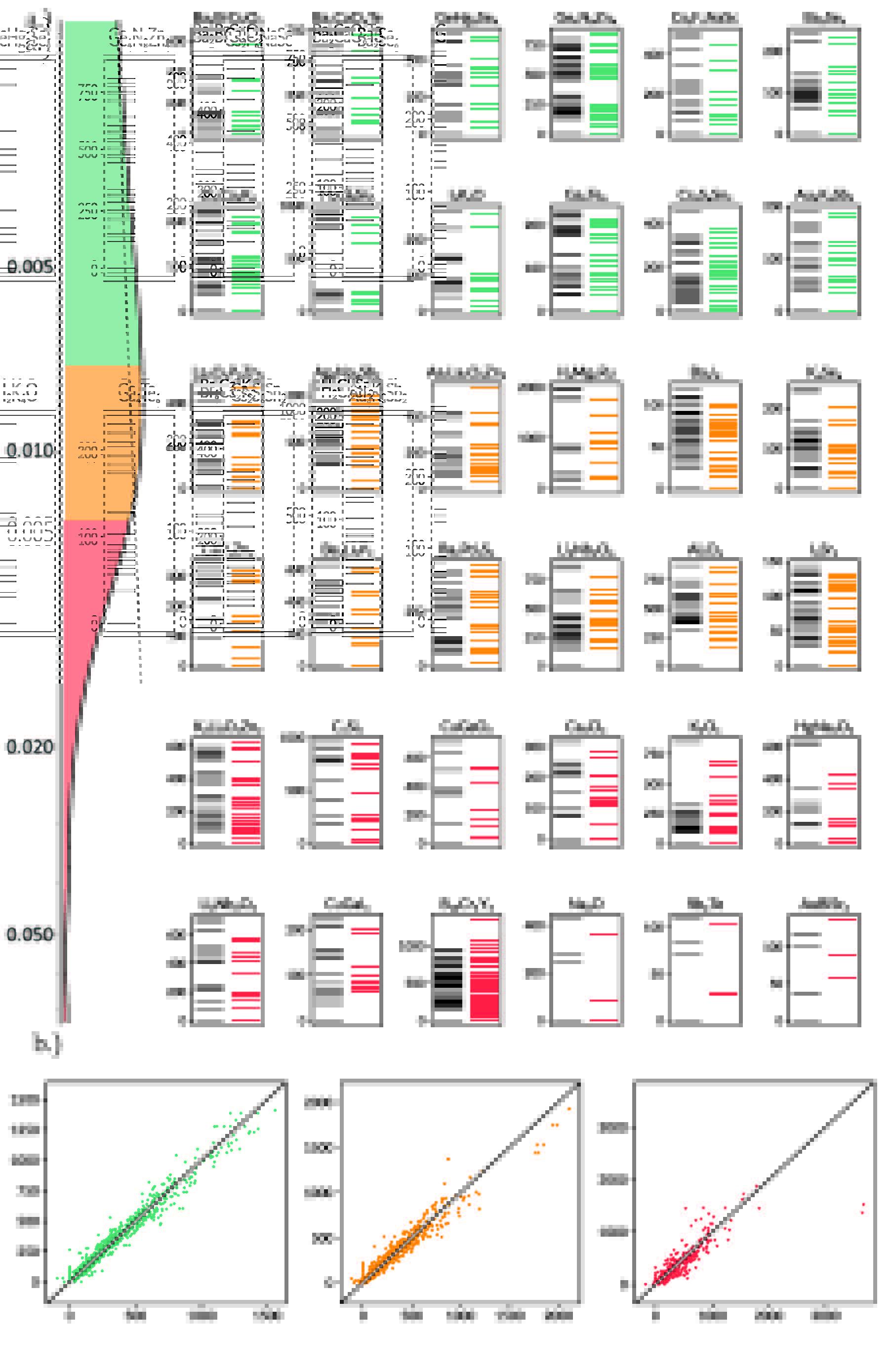}
\centering
\caption{\textbf{MVN model: Direct prediction results within the test set of the Main Database using MVN.} 
\textbf{a}. The predictions of $\Gamma$-phonon comparing with DFPT calculations (black) in 1st, 2nd and 3rd tertiles(green, yellow, and red) in [$cm^{-1}$]. The loss distribution on the left shows a peak in the 1st and 2nd tertiles.
\textbf{b}. The correlation plots (prediction Vs. ground truth) of all $\Gamma$-phonon within the test set in each tertile.  
}
\label{dynam_test_pred}
\end{figure}

\begin{figure}[H]
\includegraphics[height=0.9\textheight]{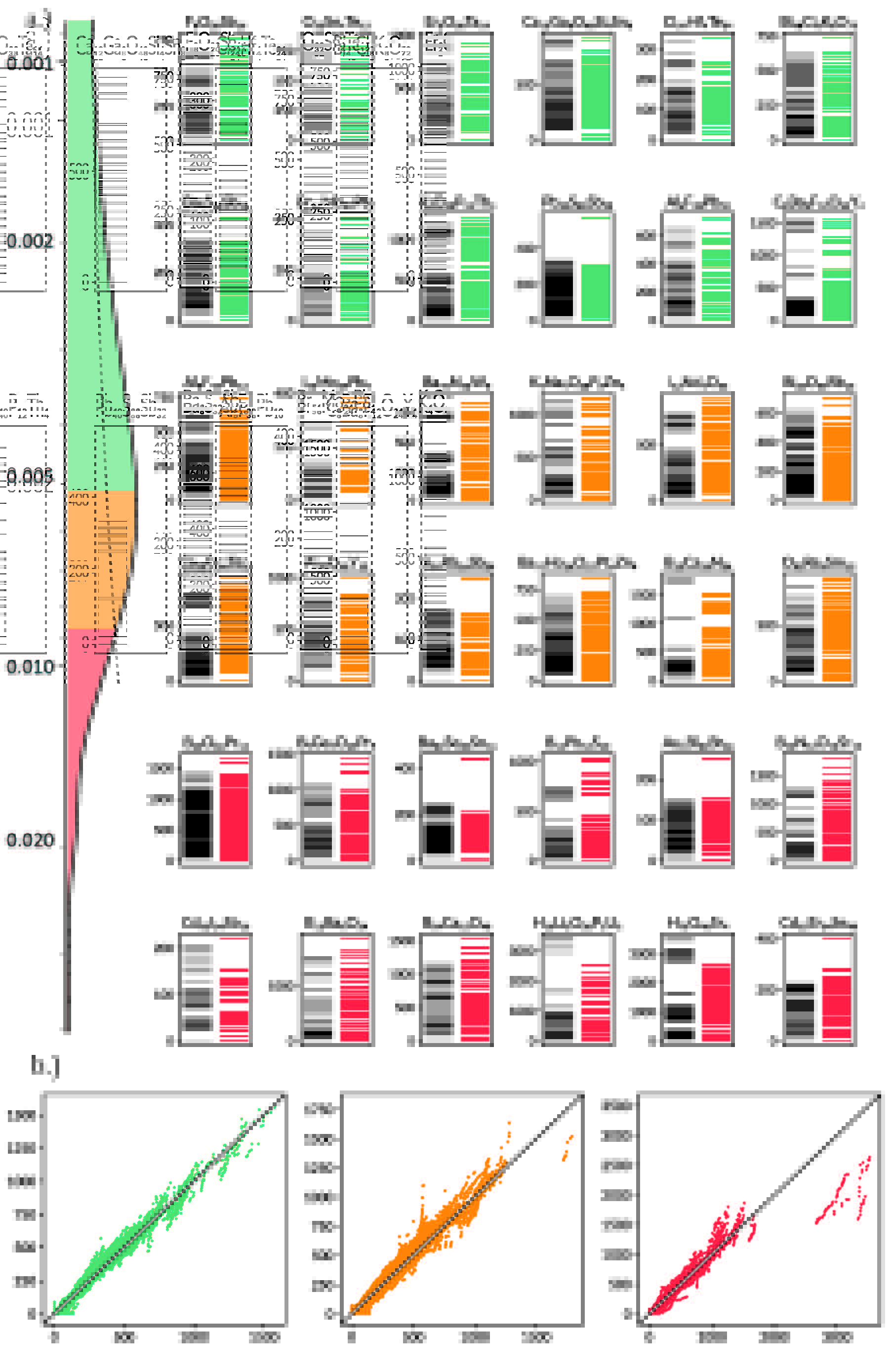}
\centering
\caption{\textbf{MVN model: Direct prediction results within test set of the Togo Database using MVN.} 
\textbf{a}. The predictions of $\Gamma$-phonon comparing with DFPT calculations (black) in 1st, 2nd and 3rd tertiles(green, yellow, and red) in [$cm^{-1}$]. The loss distribution on the left shows a peak in the 1st and 2nd tertiles.
\textbf{b}. The correlation plots (prediction Vs. ground truth) of all $\Gamma$-phonon within the test set in each tertile.  
}
\label{dynam_kyoto_pred}
\end{figure}

We present additional $\Gamma$-phonon predictions. Figure \ref{diag_train_pred}-\ref{dynam_kyoto_pred} shows the direct prediction and correlation plots of training, testing data from the Main Database, and testing data from the Togo Database using the VVN model, MVN, and k-MVN model. 

\newpage
\subsection{General Validity of Prediction on Unseen Materials}

\begin{figure}[H]
\includegraphics[height=0.9\textheight]{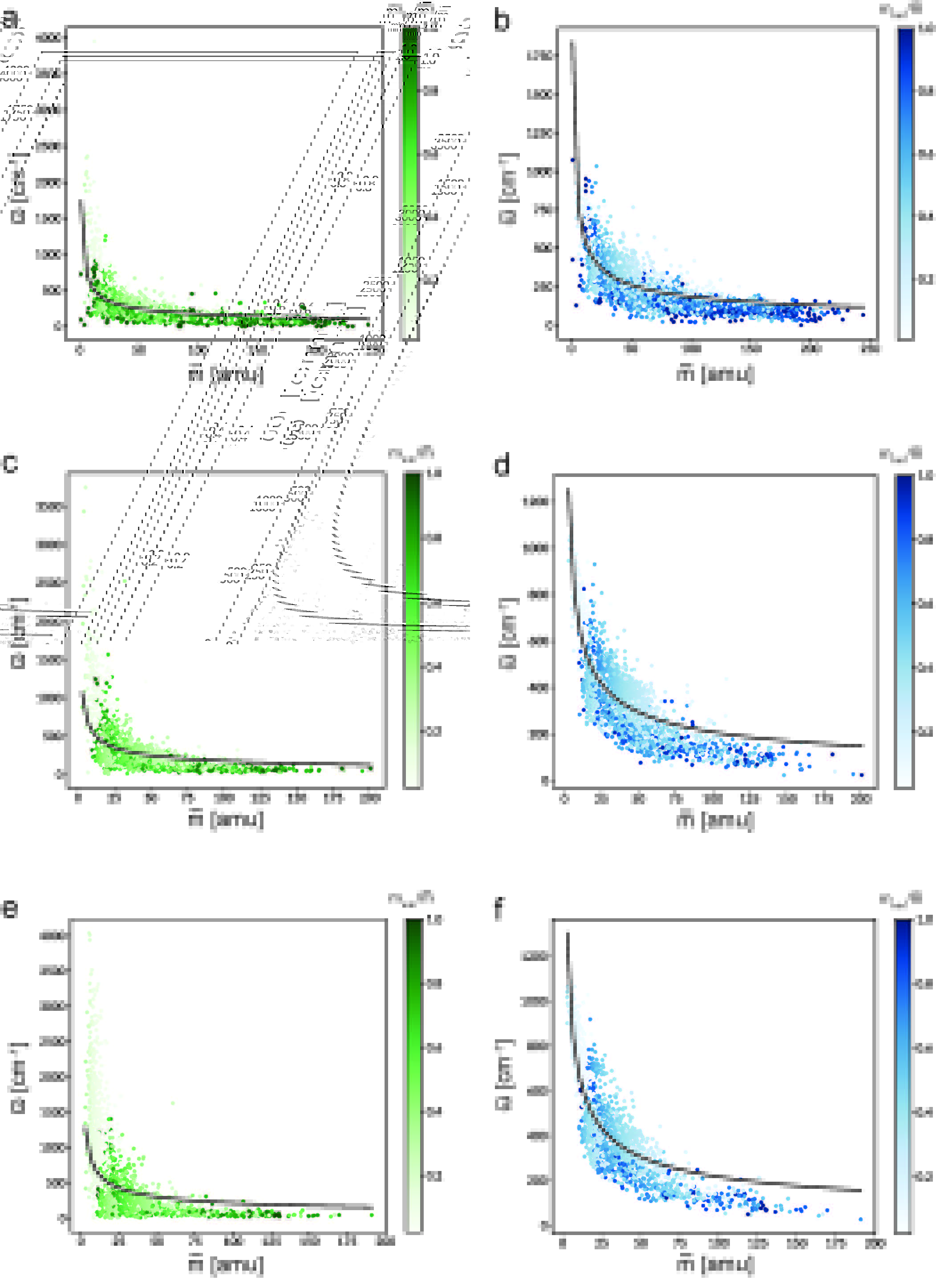}
\centering
\caption{\textbf{Evaluation of model predictions on unseen crystal structures} 
Average frequency of the predicted $\Gamma$-phonon $\bar{\omega}$ versus average atomic mass $\bar{m}$ using (\textbf{a}.,\textbf{c}.,\textbf{e}.) VVN and (\textbf{b}.,\textbf{d}.,\textbf{f}.) MVN. 5000 structures with N atoms per unit cell are randomly sampled; (\textbf{a}.,\textbf{b}.) 1$\leq N \leq 20$ (\textbf{c}.,\textbf{d}.) $21\leq N \leq 40$ (\textbf{e}.,\textbf{f}.) $41\leq N \leq 80$. The black solid lines represent the least squares that fit the hyperbolic relation $\bar{\omega} = C \bar{m}^{-1/2}$. The constant C is estimated from the fit as \textbf{a}. $C$=1743 \textbf{b}. $C$=1847 \textbf{c}. $C$=1760 \textbf{d}. $C$=2073 \textbf{e}. $C$=2113 \textbf{f}. $C$=2157. The dot colors represent the magnitudes of $m_{min}/\bar{m}$.
}
\label{mbar_omegabar_atoms}
\end{figure}

In Figure \ref{mbar_omegabar_atoms}, we plot the average phonon frequency against the quadratic mean of atomic mass in the unit cell. We apply the predictive model on 5000 unseen crystal structures from the Materials Project with atomic site number N in each unit cell (\ref{mbar_omegabar_atoms}a,b) 1$\leq N \leq 20$, (\ref{mbar_omegabar_atoms}c,d) 21$\leq N \leq 40$, (\ref{mbar_omegabar_atoms}e,f) 41$\leq N \leq 80$. For each material the average phonon frequency is related to the mass by the hyperbolic relationship $\bar{\omega} = C \bar{m}^{-1/2}$, where a constant C represents the rigidity of the crystal. The reasonable distribution of rigidity supports the physical validity of our model for unknown materials. Moreover, we characterize the non-uniformity of atomic masses in each material by computing the ratio of the minimum mass $m_{min}$ in a crystal to $\bar{m}$. The materials with high $m_{min}/\bar{m}$, containing both small and large atoms, tend to aggregate at lower $\bar{\omega}$ and higher $\bar{m}$. Whereas, the materials with low $m_{min}/\bar{m}$, containing atoms with similar mass, tend to aggregate at higher $\bar{\omega}$ and lower $\bar{m}$. These tendencies suggest that small atoms, such as hydrogen and oxygen, give high-frequency phonons which agree with the results in the previous study\cite{petretto2018high}. Note that the VVN's plots show higher and broader $\bar{\omega}$ distribution in the low $\bar{m}$ region as N gets larger, while those of MVN keep narrower distribution in the same region. This corresponds to the results in Figure \ref{diag_kyoto_pred} that the $\Gamma$-phonon predicted by VVN gets higher than the ground truth in the case the input materials are complicated. 

\subsection{Element-wise Prediction of Average Phonon Energies}

\begin{figure}[H]
\centering
\includegraphics[width=\textwidth]{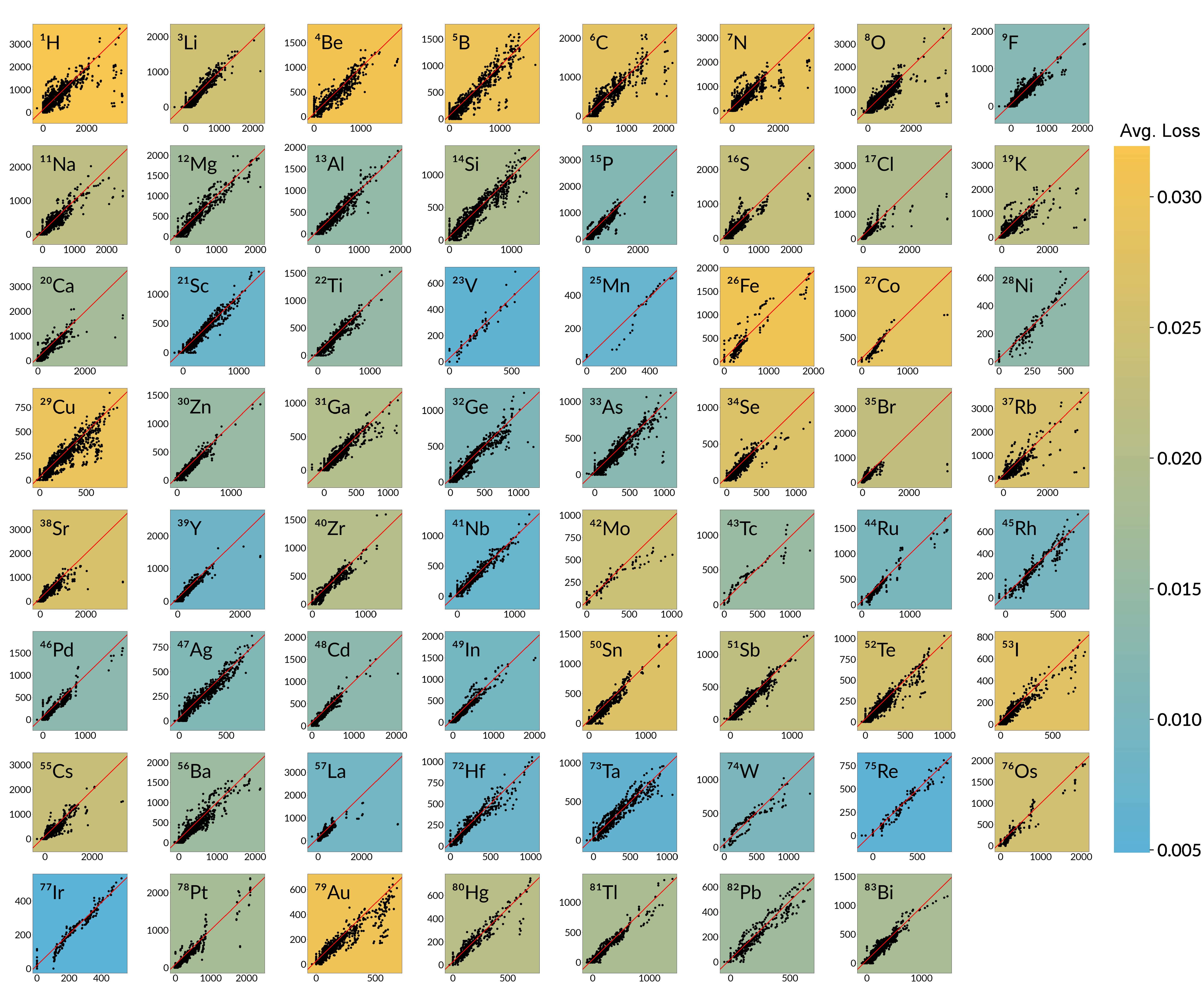}
\centering
\caption{\textbf{Element-wise correlation plot of the prediction results within train set using VVN} 
Correlation plots with 63 elements existing within the train set.  
}
\label{element_corr_diag_train}
\end{figure}

\begin{figure}[H]
\includegraphics[width=\textwidth]{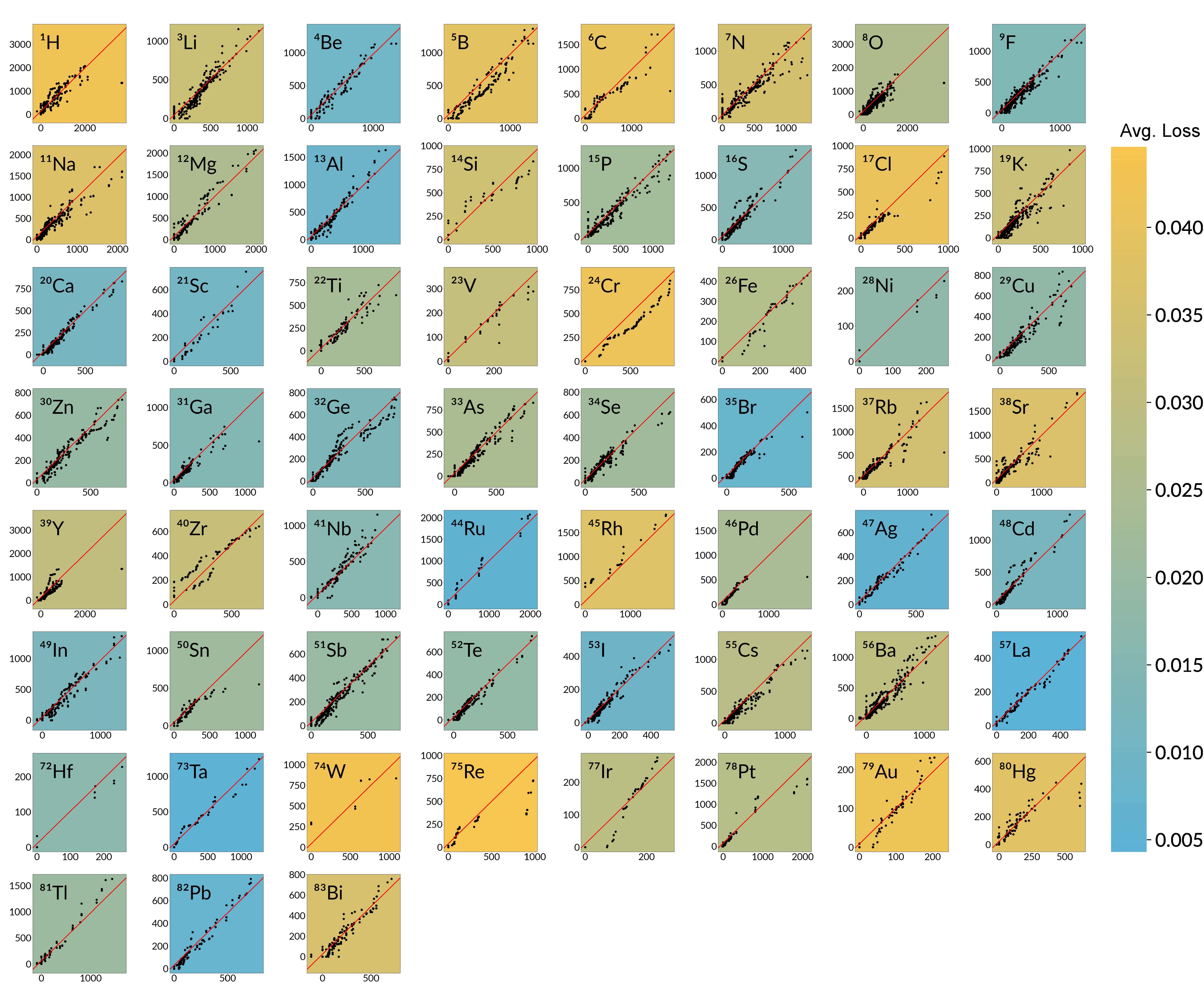}
\centering
\caption{\textbf{Element-wise correlation plot of the prediction results within test set using VVN.} 
Correlation plots with 59 elements existing within the test set.  
}
\label{element_corr_diag}
\end{figure}

\begin{figure}[H]
\includegraphics[width=\textwidth]{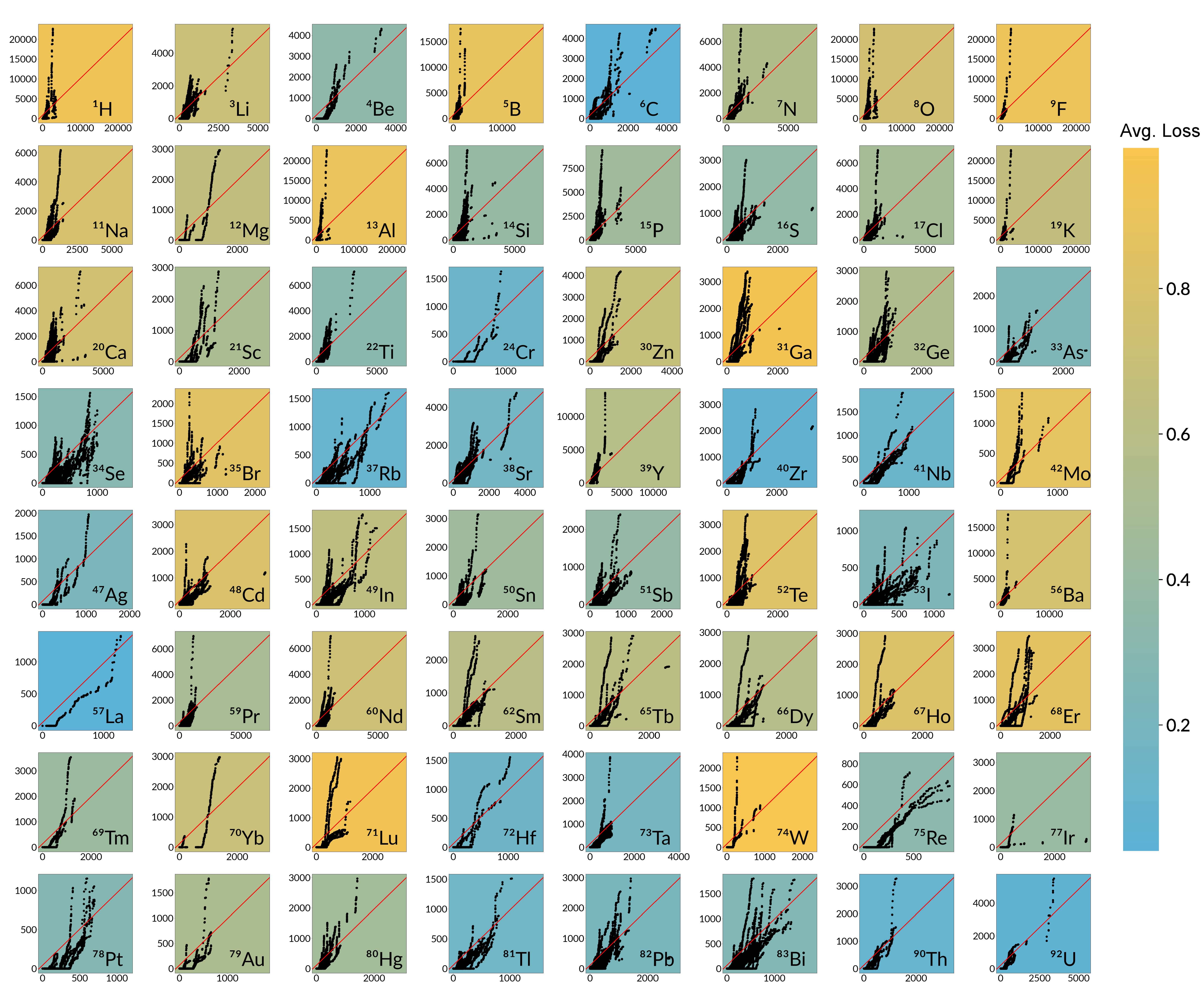}
\centering
\caption{\textbf{Element-wise correlation plot of the prediction results within train set of the Togo Database using VVN.} 
Correlation plots with 64 elements existing within the test set of complicated materials.  
}
\label{element_corr_diag_kyoto}
\end{figure}

\begin{figure}[H]
\includegraphics[width=\textwidth]{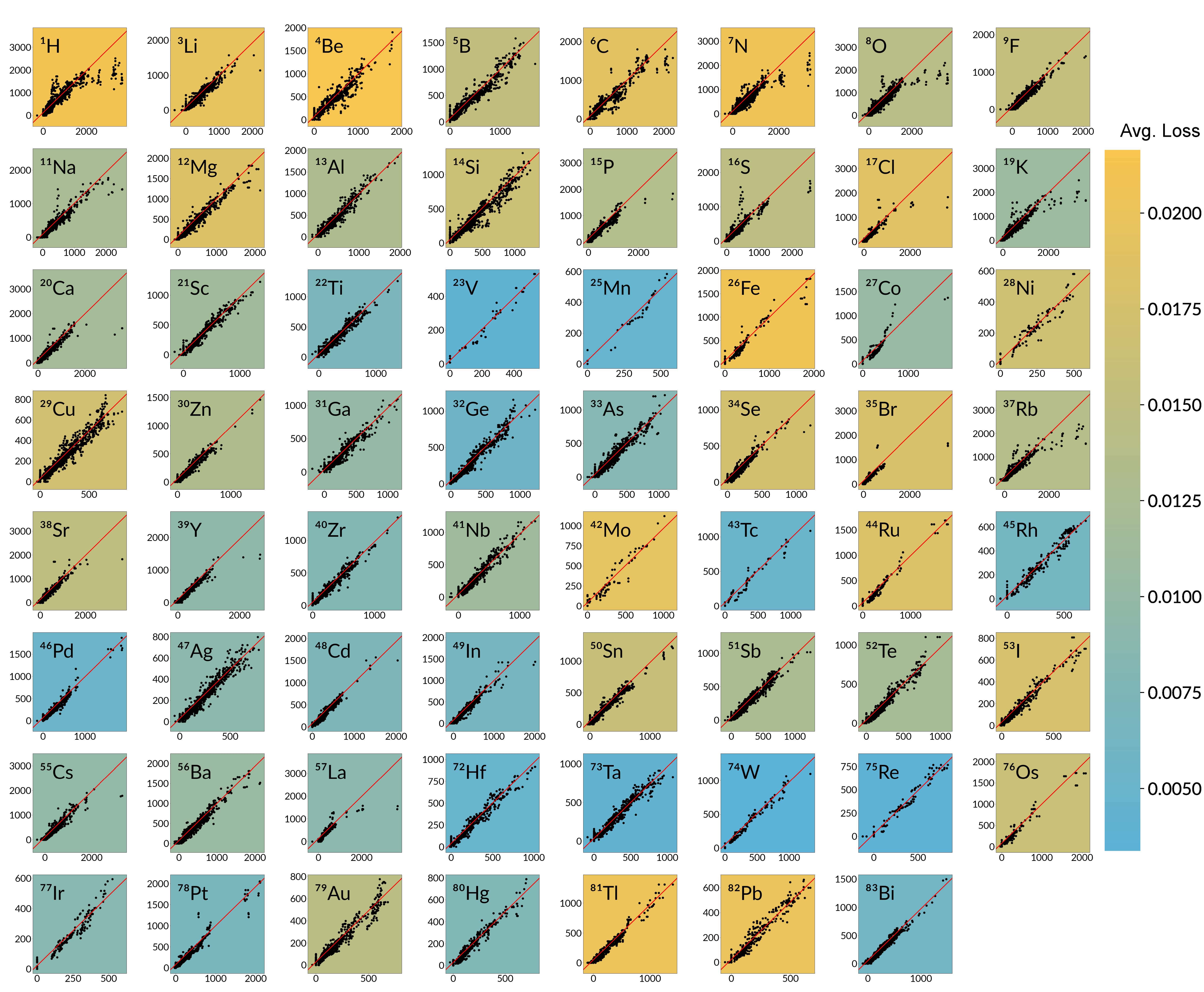}
\centering
\caption{\textbf{Element-wise correlation plot of the prediction results within train set using MVN.} 
Correlation plots with about 63 elements existing within the train set. The plotted phonon frequencies are in [$cm^{-1}$].  
}
\label{element_corr_dynam_train}
\end{figure}

\begin{figure}[H]
\includegraphics[width=\textwidth]{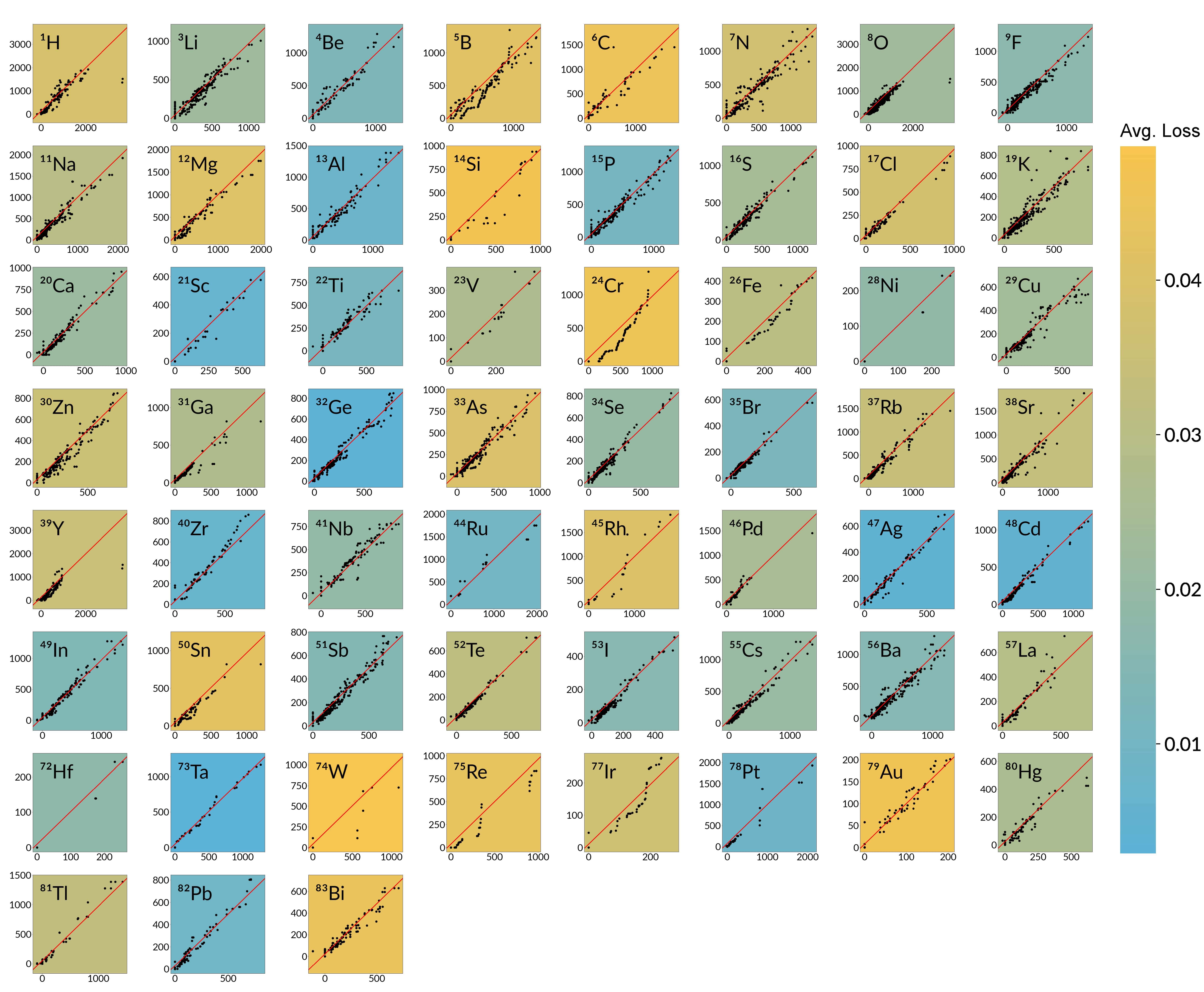}
\centering
\caption{\textbf{Element-wise correlation plot of the prediction results within test set using MVN.} 
Correlation plots with about 59 elements exist within the test set. The plotted phonon frequencies are in [$cm^{-1}$]. 
}
\label{element_corr_dynam}
\end{figure}

\begin{figure}[H]
\includegraphics[width=\textwidth]{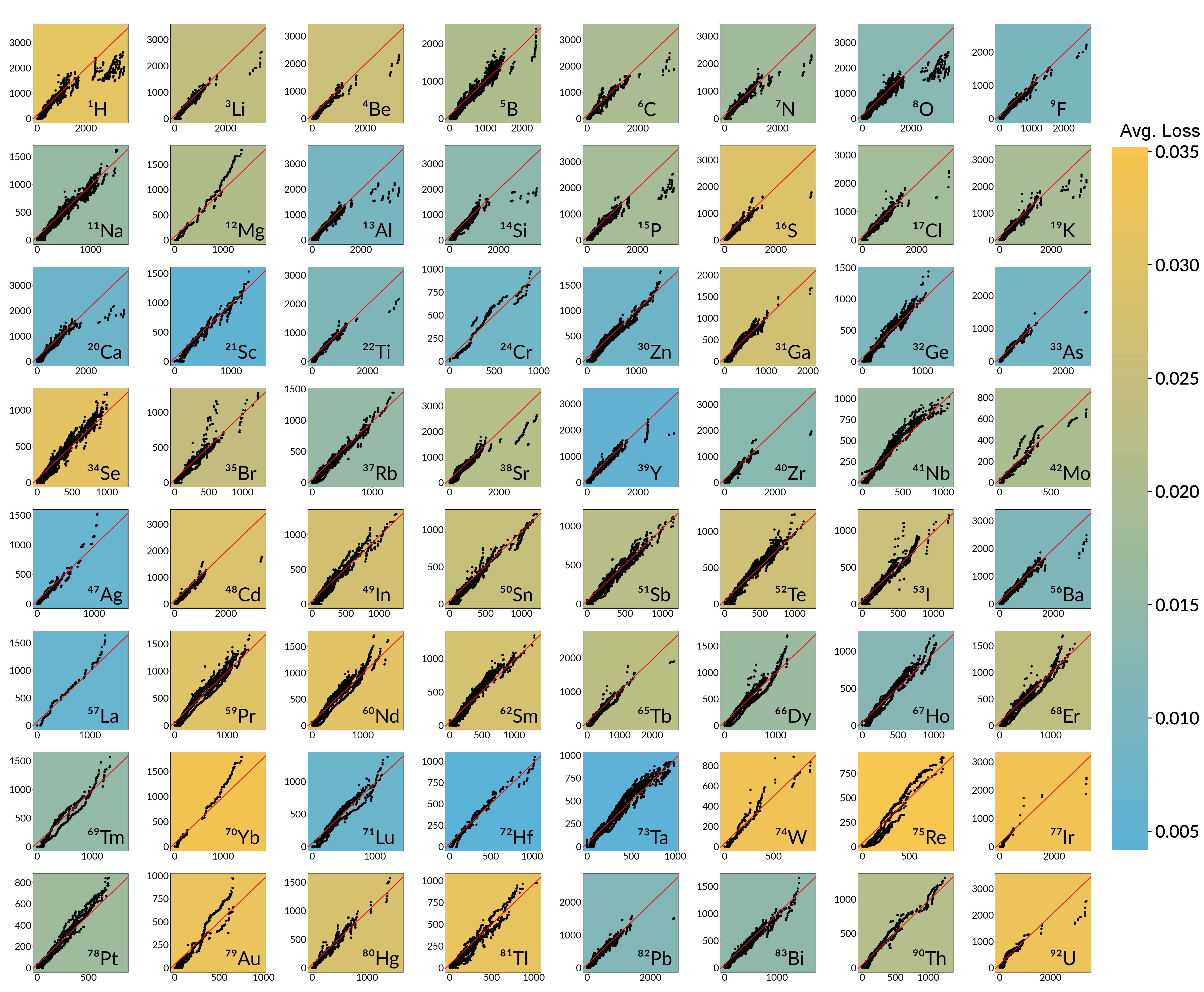}
\centering
\caption{\textbf{Element-wise correlation plot of the prediction results within test set of the Togo Database using MVN.} 
Correlation plots with about 64 elements exist within the test set of the complicated materials. The plotted phonon frequencies are in [$cm^{-1}$].  
}
\label{element_corr_dynam_kyoto}
\end{figure}

In Figure \ref{element_corr_diag_train}-\ref{element_corr_dynam_kyoto}, we illustrate the correlation plots of the $\gamma$-phonons of all element appearing in the training and testing data sets using VVN and MVN. The background color represents the value of the prediction loss (yellow to blue from low to high loss). We could see that it is not strict, but there are dependencies of the prediction accuracy on the periodicity for some parts of the elements (e.g. from $^{5}B$ to $^{9}F$). Moreover, when compared to Figure \ref{element_count_train}-\ref{element_count_kyoto}, it is generally found that elements corresponding to high prediction loss in Figure \ref{element_corr_diag_train}-\ref{element_corr_dynam_kyoto} tend to appear less frequently than those with lower errors.

\section{FULL PHONON BAND STRUCTURES PREDICTIONS}

We present more results by using the $k$-MVN model for predicting the full phonons dispersion band within the training and testing data from the Main Database and complex materials from the Togo database. Note that some of the phonons in DFPT calculation have negative values, and the model is set so that it only gives positive phonon frequency. Although the phonon bands prediction of some materials does not match up exactly with the DFPT calculation, some features of the bands, such as an average frequency and band gaps, are well captured by the model.

\begin{figure}[H]
\includegraphics[width=\textwidth]{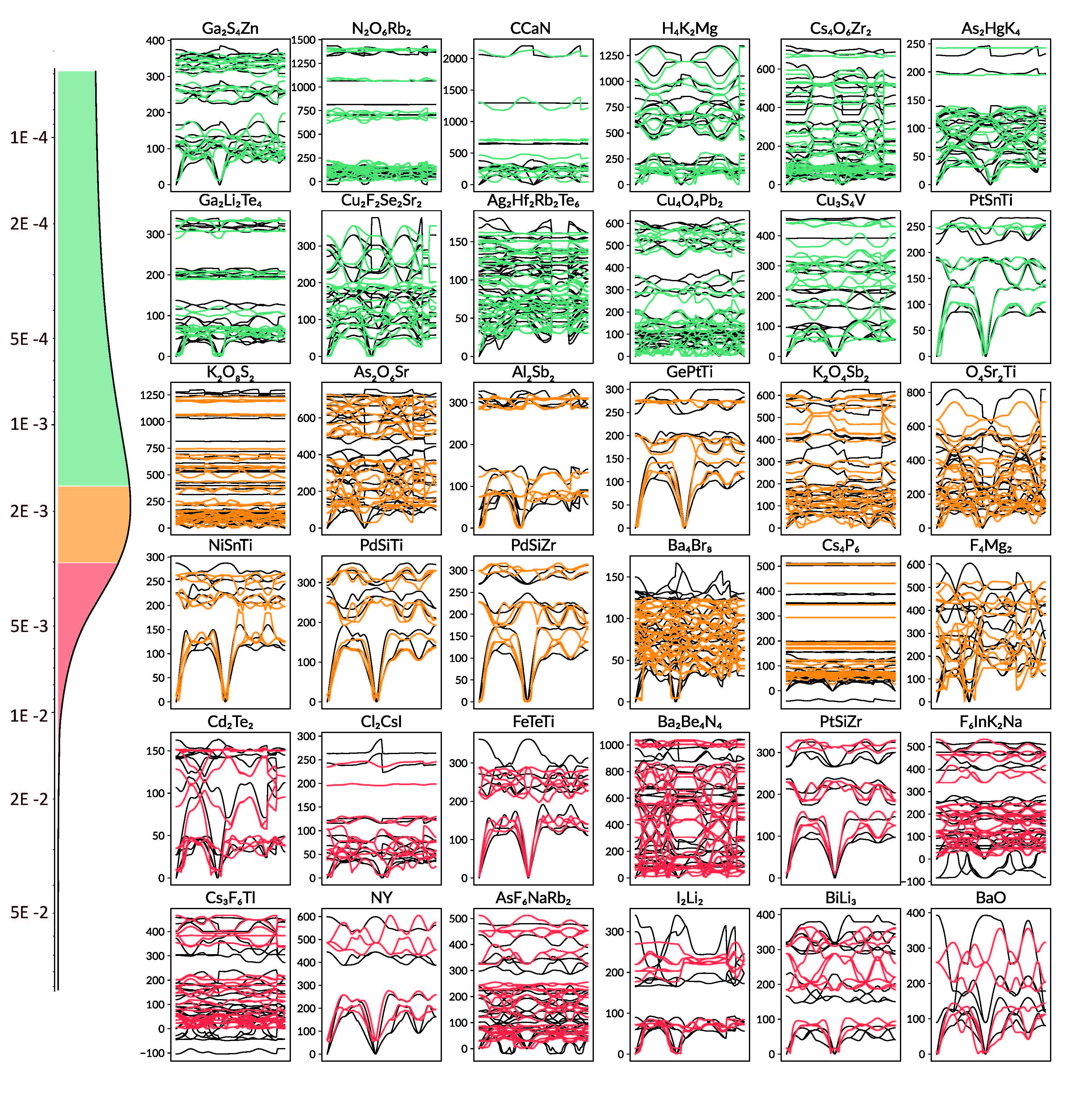}
\centering
\caption{\textbf{Direct prediction result of phonon band structures within the training set of the Main Database using $k$-MVN.} 
Each band contains 100 k-points across the Brillouin zone which varies for each material depending on its structure. The band calculated from DFPT are labeled in black, and green, yellow, and red lines represent the predicted phonon bands in 1st-3rd tertiles.
}
\label{kmvn_band_train}
\end{figure}

\begin{figure}[H]
\includegraphics[width=\textwidth]{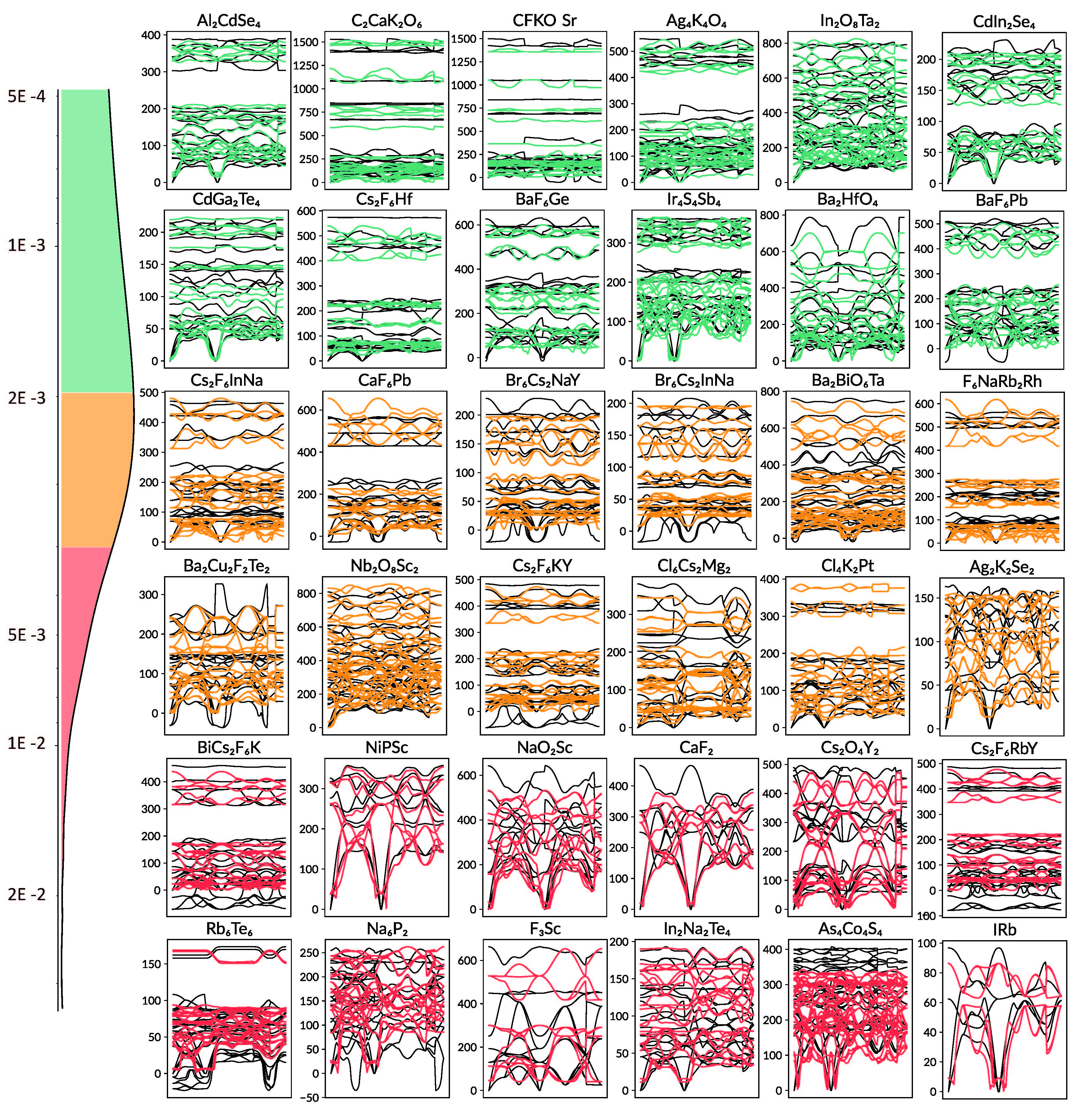}
\centering
\caption{\textbf{Direct prediction result of phonon band structures within the testing set of the Main Database using $k$-MVN.} 
Each band contains 100 k-points across the Brillouin zone which varies for each material depending on its structure. The band calculated from DFPT are labeled in black, and green, yellow, and red lines represent the predicted phonon bands in 1st-3rd tertiles.
}
\label{kmvn_band_test}
\end{figure}

\begin{figure}[H]
\includegraphics[width=\textwidth]{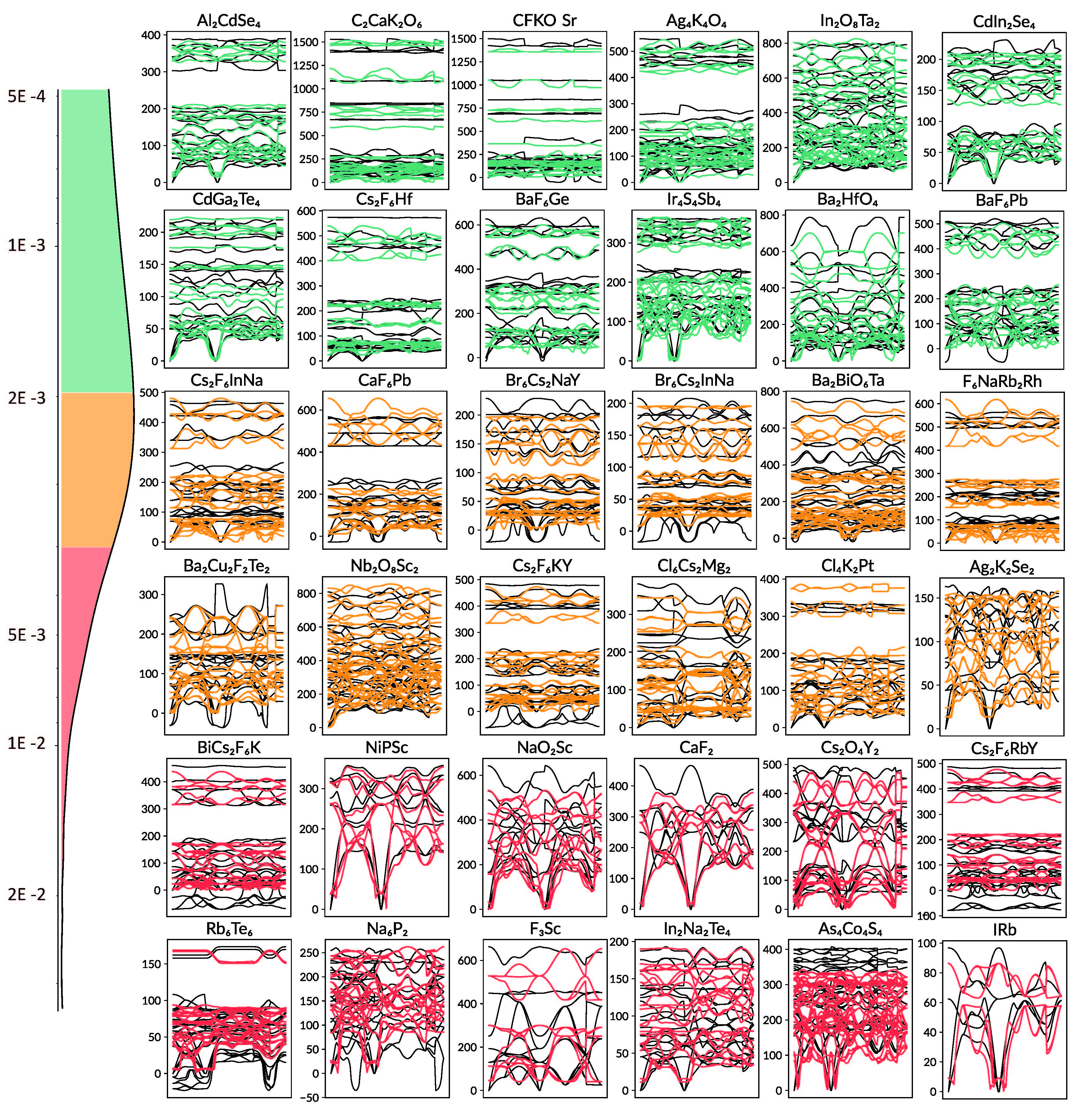}
\centering
\caption{\textbf{Direct prediction result of phonon band structures of the Togo Database using $k$-MVN.} 
Each band contains 100 k-points across the Brillouin zone which varies for each material depending on its structure. The band calculated from DFPT are labeled in black, and green, yellow, and red lines represent the predicted phonon bands in 1st-3rd tertiles.
}
\label{kmvn_band_kyoto}
\end{figure}

Figure \ref{kmvn_band_train}-\ref{kmvn_band_kyoto} provide more results of the full phonon bands prediction using $k$-dependent matrix virtual nodes model.

\newpage
\section{Applications}
\subsection{Phonon Prediction on a Topological Weyl Semimetal}
The prediction of LaAlGe $\Gamma$-phonon obtained from $k$-MVN model compared to the experimental results \cite{Thanh2020Kohn}. The phonon dispersion of LaAlGe- a Type-II Weyl semimetal, was measured by inelastic neutron and X-ray techniques along the high symmetric points and Weyl points.

\begin{figure}[h!]
\includegraphics[width=\textwidth]{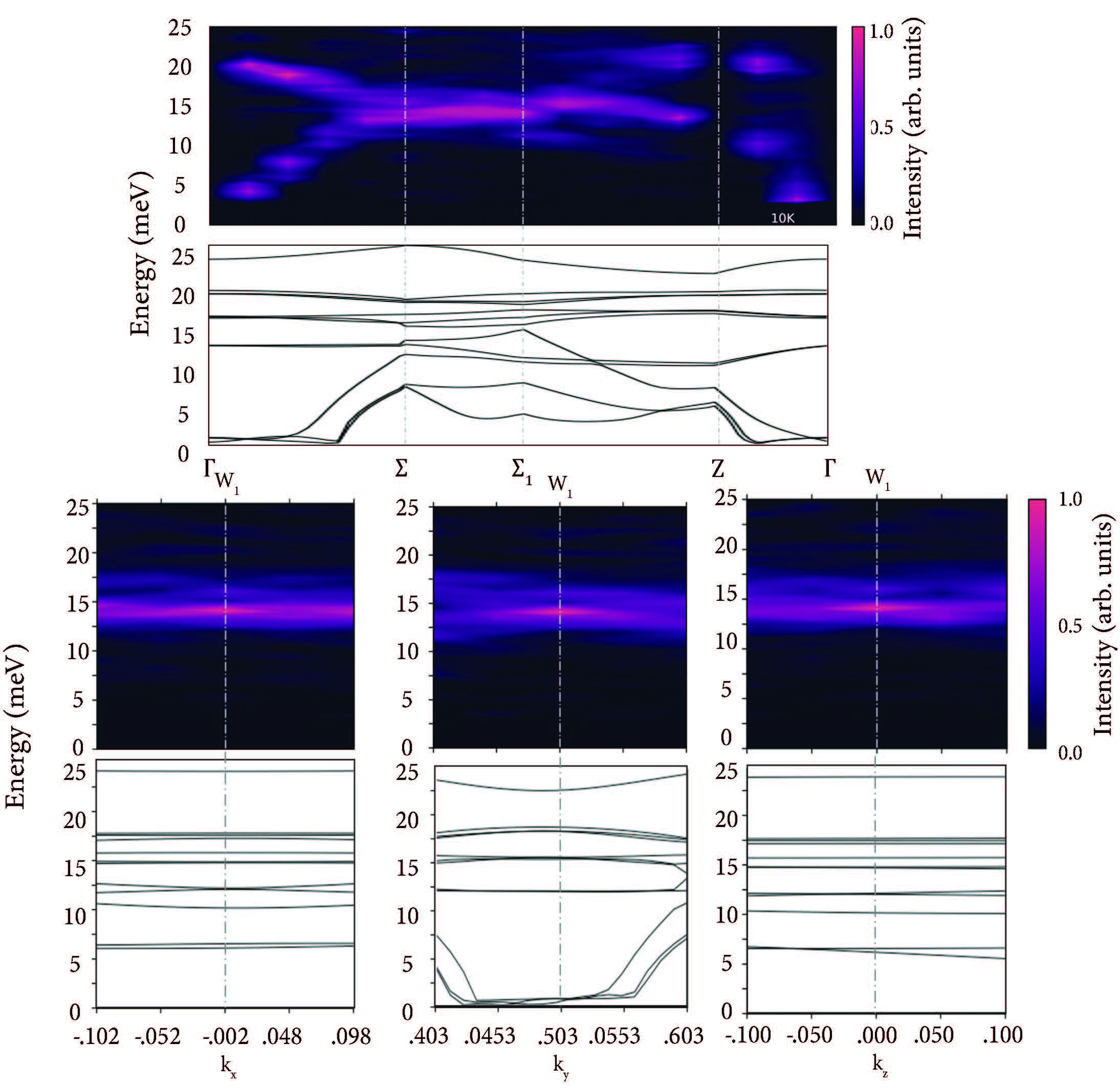}
\centering
\caption{\textbf{Comparison of LaAlGe's phonon spectra acquired with experiments and machine learning} 
\textbf{(Top)} LaAlGe phonon bands along the highly symmetric points. \textbf{(Bottom)} The phonon band structures along the x, y, z directions within the Brillouin zone near one of the Weyl points W1. The plotted phonon frequencies are in [meV]. 
}
\label{exp_laalge}
\end{figure}

\subsection{Phonon Prediction in Alloy Systems}
\begin{figure}[h!]
\includegraphics[width=\textwidth]{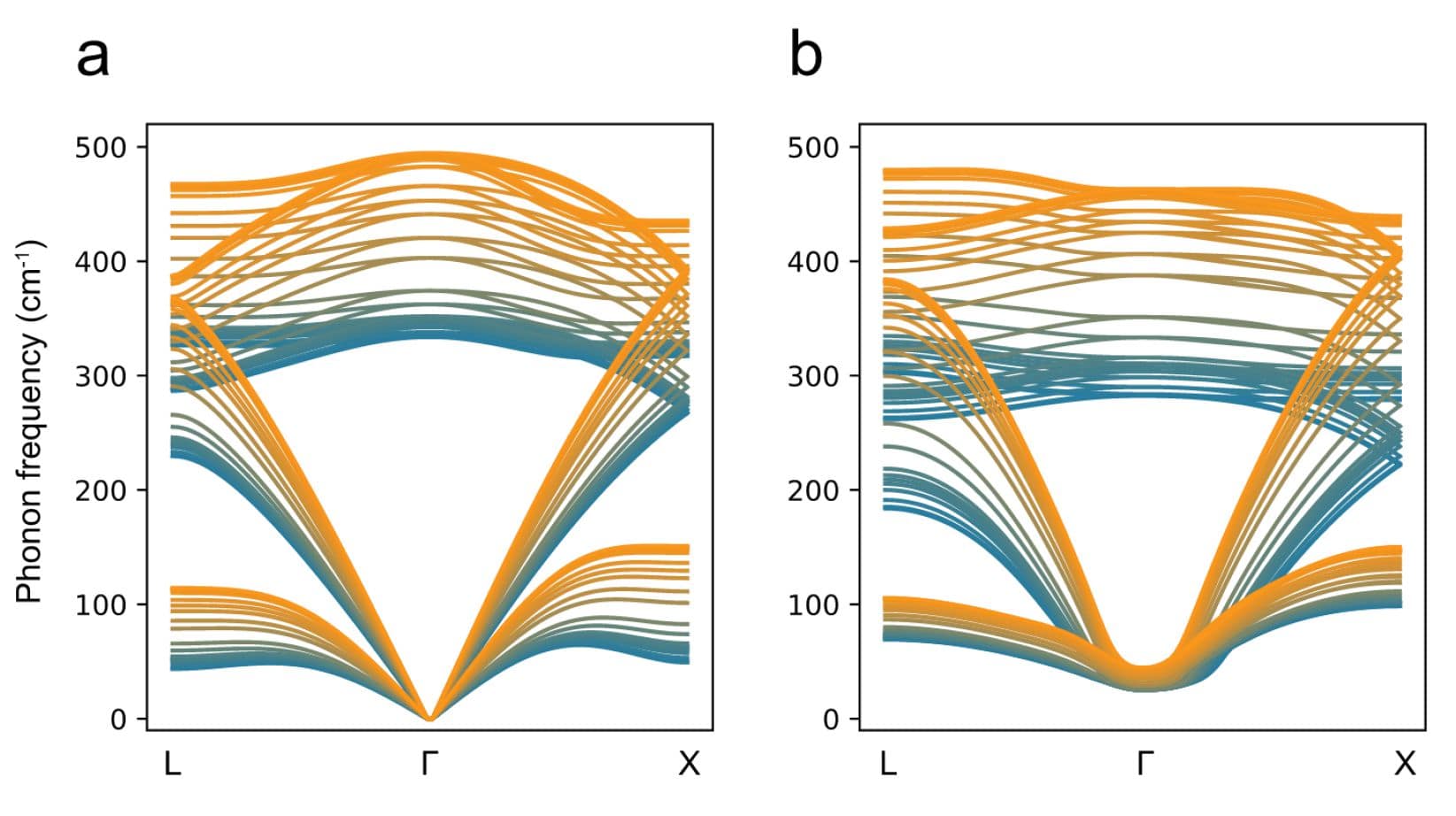}
\centering
\caption{\textbf{Phonon spectra of SiGe alloy with various compositions.} 
Yellow and blue indicate pure Si and Ge, respectively, and the intermediate colors represent alloys of different composition ratios.  We demonstrated phonon prediction of $\text{Si}_p\text{Ge}_{1-p}$ with p = 0.00, 0.01, 0.05, 0.10, 0.13, 0.15, 0.17, 0.20, 0.30, 0.40, 0.60, 0.70, 0.80, 0.85, 0.90, 0.96, 0.98, 0.99, 0.995, and 1.00 (from blue to orange). We show phonon along L, $\Gamma$, and X points. \textbf{a}. Ground truth. \textbf{b}. Prediction.
}
\label{sige_qichen}
\end{figure}

One of the most important applications of our prediction model is phonon bands of alloy systems. Crystalline alloys have disorders whose atomic positions can be periodic. However, the system is not periodic because atomic elements' distribution at one site is probabilistic. Therefore, we need to consider the atomic compositions when we embed each site's atomic mass, as the virtual-crystal approximation (VCA)\cite{bellaiche2000virtual} is based on. For example, given a binary alloy with composition $V_pW_{1–p}(0 \leq p \leq 1)$, the input alloy encoding vector $\mathscr{Z}$ can take the following form
\begin{equation}
 \mathscr{Z} = [0, ... ,pm_V, ... , (1 - p)m_W, ... , 0]
\label{eq_alloyZ}
\end{equation}
where the two-hot encoding $pm_V$ and $(1 - p)m_W$ are located at the vector indices corresponding to the atomic numbers of $V$ and $W$, respectively, weighted by composition. With this definition of equation (\ref{eq_alloyZ}), the embedded feature can be directly reduced to pure phase one-hot encoding V (or W) by simply setting p = 1 (or p = 0), and it can be generalized to more complicated alloys directly.  \\
We demonstrate the power of this approach with the binary alloy of SiGe. We always assume SiGe alloy has eight atoms per cubic unit cell with a side of 5.466 \AA, which is the intermediate of the crystalline silicon and germanium\cite{jain2013commentary}. As the input of kMVN model, we convert the unit cell into a primitive cell with two atoms. As the ground-truth label, we used the phonon bands calculated with VCA approximation. \\
Figure.\ref{sige_qichen} shows the result of the phonon band prediction of SiGe alloy. Both ground truth (Figure.\ref{sige_qichen}a) and prediction (Figure.\ref{sige_qichen}b) shows similar shapes of the band structures along the highly-symmetric points. Yellow and blue colors depict the composition ratios of silicon and germanium, respectively. The lighter atoms generally offer phonons of higher frequencies. Our result follows the principle, as the alloys give higher phonon frequencies as the compositions of silicon get larger. 

\begin{figure}[h!]
\includegraphics[width=\textwidth]{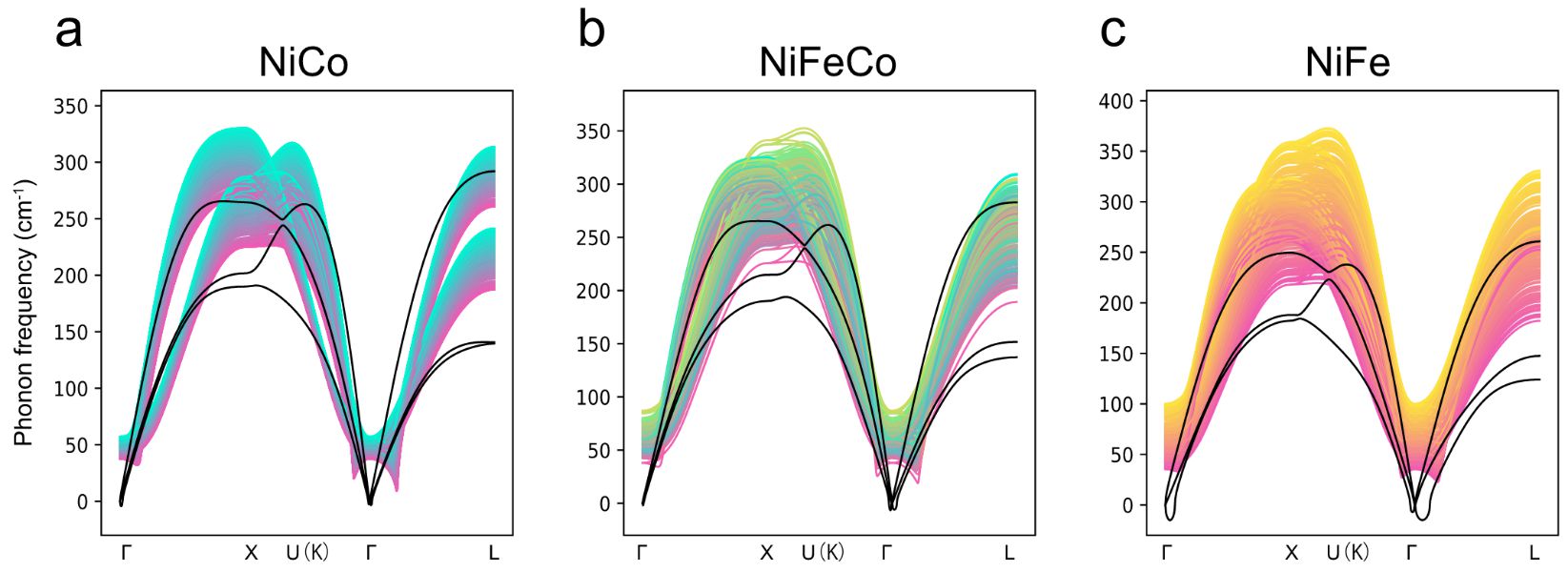}
\centering
\caption{\textbf{Phonon spectra of high entropy alloys with various compositions.} 
Red, blue and yellow indicate pure Ni, Co, and Fe, respectively, and the intermediate colors represent alloys of different composition ratios. The black lines in each figure are ground truth computed by DFPT. \textbf{a}. NiCo. \textbf{b}. NiFeCo. \textbf{c}. NiFe.
}
\label{heas_sai}
\end{figure}

We next show the phonon prediction of binary alloys NiCo, NiFe, and a ternary alloy NiFeCo. For each material, we generated the random compositions 200 times and observed how the phonon prediction went. As the ground-truth label (black lines in each of Figure \ref{heas_sai} ), we used the phonon bands calculated with DFPT approximation. As the input of the DFPT calculation, we used supercells that contain 64 atoms for NiCo and NiFe and 108 atoms for NiFeCo, respectively.  Figure \ref{heas_sai} shows the predicted phonon of the three alloys. The combination of red, blue, and yellow colors represent the compositions Ni, Co, and Fe, respectively. The spectra were arranged from left to right in the ascending order of the prediction loss: (a) NiCo, (b) NiFeCo, and (c) NiFe. This tendency can be explained by the atomic masses of the alloy's components. Ni and Co have almost the same atomic mass (58.69u and 58.93u), and Fe has a lower value (55.85 u). Compared to NiFe, NiCo and NiFeCo have smaller differences in the atomic masses of the elements they have, which makes the atomic mass encoding more stable. \\
 
\begin{figure}[h!]
\includegraphics[width=\textwidth]{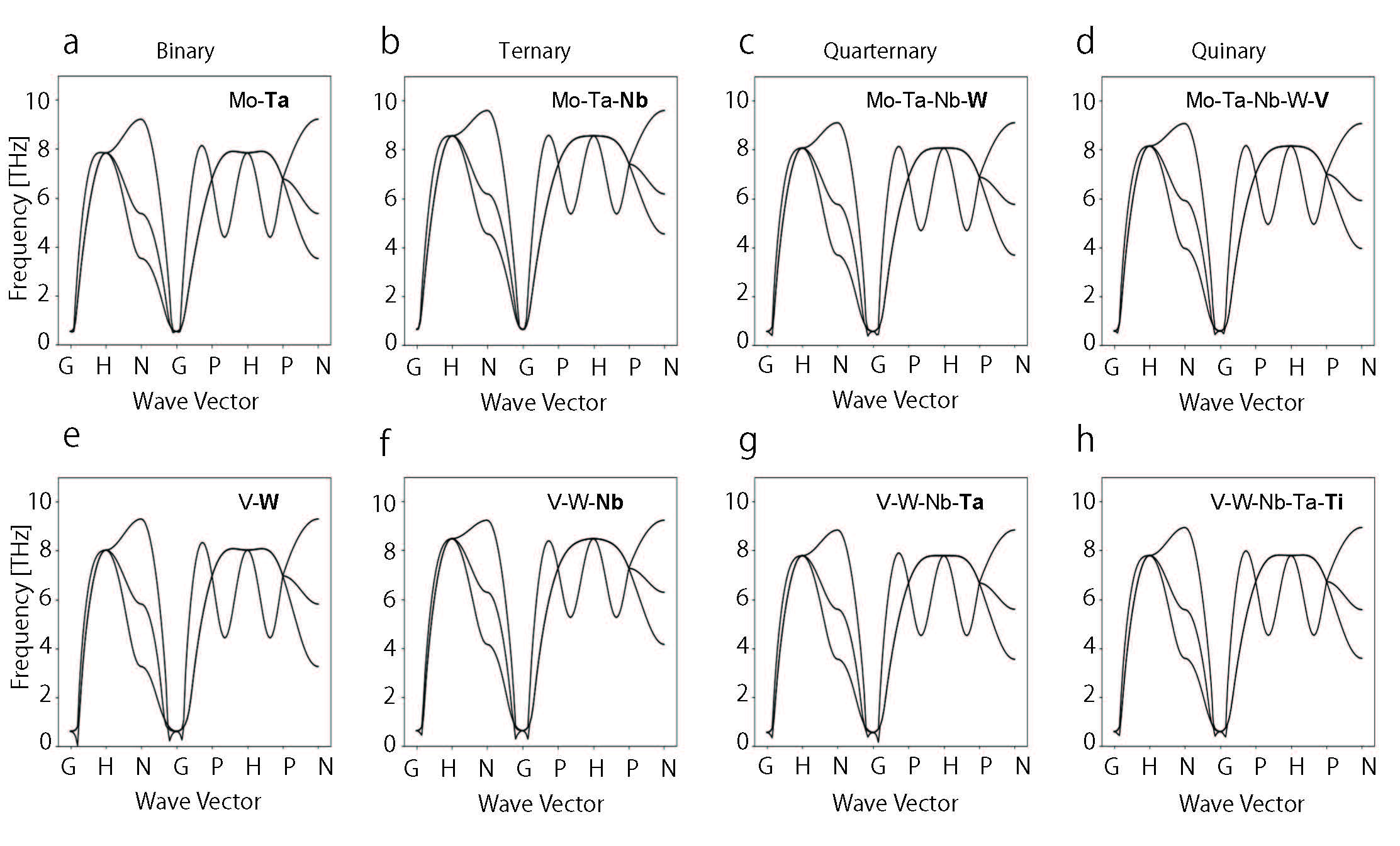}
\centering
\caption{\textbf{Phonon spectra from binary to 5-component high entropy alloys} 
Predicted phonon spectra with increasing constituent elements: From binaries to 5-component high entropy alloys. The added element for each alloy from left to right is shown in bold. The plotted phonon frequencies are in [THz]. 
\textbf{a}. MoTa.
\textbf{b}. MoTaNb.
\textbf{c}. MoTaNbW.
\textbf{d}. MoTaNbWV.
\textbf{e}. VW.
\textbf{f}. VWNb.
\textbf{g}. VWNbTa.
\textbf{h}. VWNbTaTi.
}
\label{hea_broadening}
\end{figure}

We have shown that our $k$-MVN model could handle phonon prediction of binary and ternary alloys. Here we demonstrate even higher component alloys. Here, we start with binaries, moving to ternaries, then quaternaries, and finally to the five-component quinaries. Figure.\ref{hea_broadening} shows the  predicted spectra of MoTaNbWV and VWNbTaTi, as well as their lower component alloys. \\
In high entropy alloy systems, both force constants and masses have fluctuated. Therefore, we need to average these properties to represent the disordered states as primitive cells or supercells. In Figure.\ref{hea_broadening}a-d, the number of MoTaNbWV's components increased from 2 to 5. We show a similar property for the case of VWNbTaTi, in Figure.\ref{hea_broadening}e-h. Our prediction follows atomic mass embedding of equation (\ref{eq_alloyZ}), which weights the mass of the components with the configuration ratios. Even without incorporating the force constant averaging, our model could generate phonon spectra similar to the simulation results with special quasirandom structures (SQS)\cite{kormann2017phonon}. 
We show a similar property for the case of VWNbTaTi, in Figure.\ref{hea_broadening}e-h. \\

\section{$\Gamma$-PHONON DATABASE}
In previous work, the full phonon band structures and derived quantities for 1521 semiconducting inorganic crystals are presented\cite{petretto2018high}. Given the high-quality phonon prediction using the machine-learning approach, particularly the MVN approach for complex materials, here we present a new phonon database containing the phonon spectra for the entire 146,323 Materials in Materials Project (MP) as of 2022 computed by the MVN approach. Given the importance of zone-center $\Gamma$-phonons, which are measurable with more accessible equipment like Raman scattering, here we limit the database to $\Gamma$-phonons and will leave the full phonon database for future works. This new $\Gamma$-phonon database offers the possibility to analyze the lattice dynamics for many compounds. It could be used as a useful tool to be compared with Raman scattering results for crystalline materials. The $\Gamma$-phonon database generated with the MVN method is available at \url{https://osf.io/k5utb/}.   \\

\begin{figure}[h!]
\includegraphics[width=\textwidth]{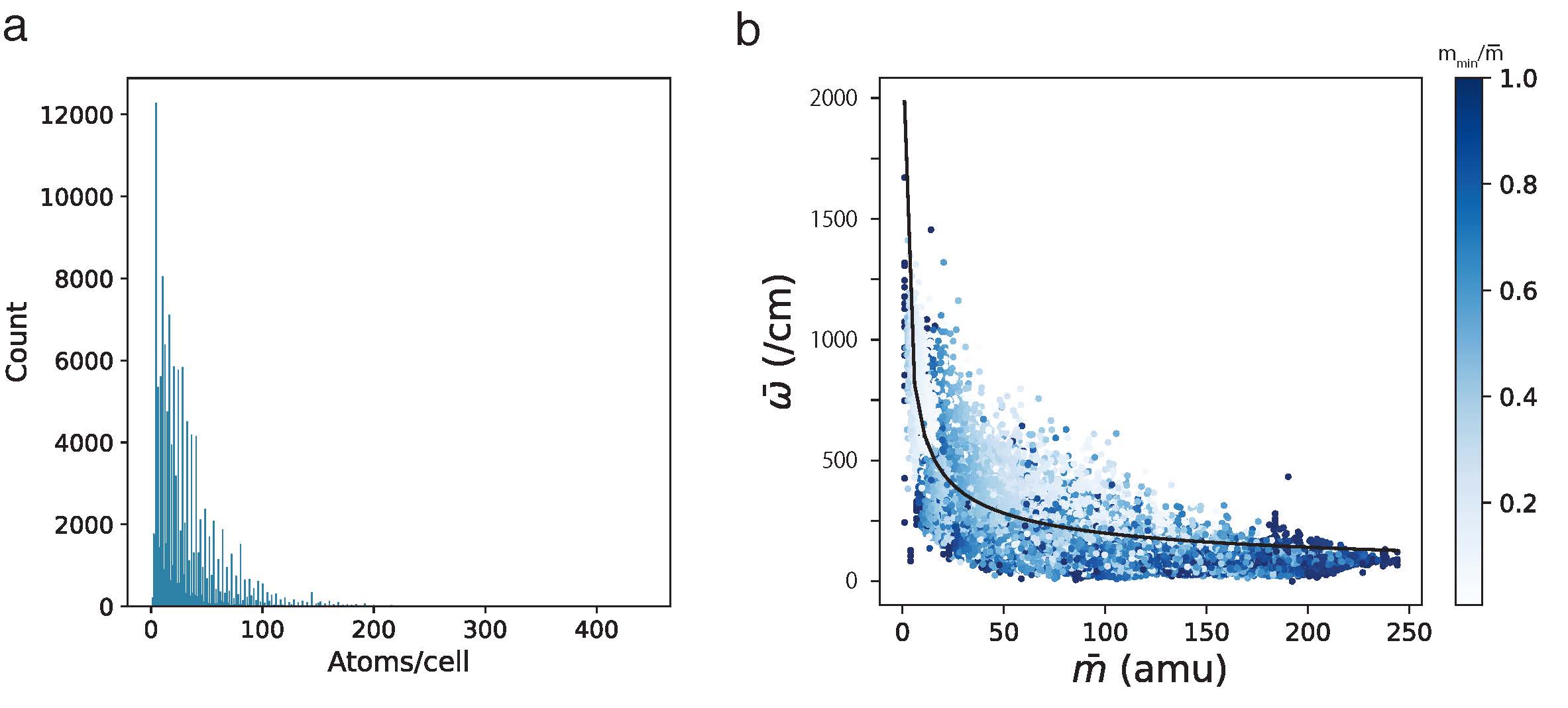}    
\centering
\caption{\textbf{The profile of our $\Gamma$-phonon database.} 
\textbf{a}. The number of atoms per unit cell in the dataset. The dataset contains 146323, and the number of atoms per unit cell ranges from 1 to 444. \textbf{b}. Average frequency of the predicted $\Gamma$-phonon $\bar{\omega}$ versus average atomic mass $\bar{m}$. The colors of the dots represent the magnitude of $m_{min}/\bar{m}$. The black solid lines represent the least squares that fit the hyperbolic relation $\bar{\omega} = C \bar{m}^{-1/2}$. The constant C is estimated from the fit as 1993. The colors of the dots represent the ratio $m_{min}/\bar{m}$.
}
\label{hist_gdb}
\end{figure}

The whole database is stored as a dictionary, whose keys are the MP ID number. The dictionary values for each material are also dictionaries, which contain basic information such as the number of atoms per unit cell, chemical formula, and others. We show the values of each dictionary in Table \ref{tab_dictionary}. We provide the test from the mass-spring model on the validity of the results, which are summarized in Figure \ref{hist_gdb}. The data displays a spread around the hyperbolic fit since the phonon frequencies are the outcome of the interplay of the whole set of interatomic force constants and the different masses of the elements composing the material. It can be noticed some trends can be recognized with respect to the masses of the components. Systems with non-uniform masses (identified by the small ratio $m_{min}/\bar{m}$), tend to lay on a different hyperbolic curve with respect to more uniformly weighted systems.
\\
\begin{table}[h!]
\begin{center}
\caption{The components of the database for each material}\label{tab_dictionary}%
\begin{tabular}{ccc}
Key & Data type & Description  \\
\hline\hline
material\_id & string & MP ID number \\
nsites & integer & The number of atoms per unit cell\\
formula & string & Chemical formula \\
space\_group & integer & The index of the space group \\
structure & string & The material structures of the CIF format \\
spectra & list & The values of $\Gamma$-phonon frequencies [\text{$cm^{-1}$}]  \\
\hline
\end{tabular}
\end{center}
\end{table}

\clearpage

\nocitesupp{*}
\bibliographystylesupp{abbrv}
\bibliographysupp{supplement}